\documentclass[fleqn,usenatbib]{mnras}

\usepackage{newtxtext,newtxmath}

\usepackage[T1]{fontenc}

\DeclareRobustCommand{\VAN}[3]{#2}
\let\VANthebibliography\thebibliography
\def\thebibliography{\DeclareRobustCommand{\VAN}[3]{##3}\VANthebibliography}

\usepackage{graphicx}	
\usepackage{amsmath}	
\usepackage{multirow}
\usepackage{tabularx,booktabs}

\usepackage[table,dvipsnames]{xcolor}

\newcommand{\hi}{H{\sc i}}

\title[Cosmological constraints from low redshift 21\,cm intensity mapping with machine learning]{Cosmological constraints from low redshift 21\,cm intensity mapping with machine learning}

\author[C. P. Novaes]{Camila P. Novaes$^{1}$\thanks{E-mail: camilapnovaes@gmail.com (CPN)}, 
Eduardo J. de Mericia$^{1}$, %
Filipe B. Abdalla$^{1,2,3,4}$, %
Carlos A. Wuensche$^{1}$, %
\newauthor 
Larissa Santos$^{5}$, %
Jacques Delabrouille$^{6,7,8,9}$, %
Mathieu Remazeilles$^{10}$, %
and Vincenzo Liccardo$^{1}$
\\
$^{1}$Instituto Nacional de Pesquisas Espaciais, Av. dos Astronautas 1758, Jardim da Granja, S\~ao Jos\'e dos Campos, SP, Brazil\\
$^{2}$University College London, Gower Street, London, WC1E 6BT, UK\\
$^{3}$Instituto de F\'{i}sica, Universidade de S\~ao Paulo, R. do Mat\~ao, 1371 - Butant\~a, 05508-09 - S\~ao Paulo, SP, Brazil\\
$^{4}$Department of Physics and Electronics, Rhodes University, PO Box 94, Grahamstown, 6140, South Africa\\
$^{5}$Center for Gravitation and Cosmology, College of Physical Science and Technology, Yangzhou University, 225009, China\\
$^{6}$CNRS-UCB International Research Laboratory, Centre Pierre Bin\'etruy, IRL2007, CPB-IN2P3, Berkeley, USA\\
$^{7}$Laboratoire Astroparticule et Cosmologie (APC), CNRS/IN2P3, Universit\'e Paris Diderot, 75205 Paris Cedex 13, France\\
$^{8}$IRFU, CEA, Universit\'e Paris-Saclay, 91191 Gif-sur-Yvette, France\\
$^{9}$Department of Astronomy, School of Physical Sciences, University of Science and Technology of China, Hefei, Anhui 230026\\
$^{10}$Instituto de Física de Cantabria (CSIC-Universidad de Cantabria), Avda. de los Castros s/n, E-39005 Santander, Spain
}

\date{Accepted XXX. Received YYY; in original form ZZZ}

\pubyear{2022}

\begin{document}
\label{firstpage}
\pagerange{\pageref{firstpage}--\pageref{lastpage}}
\maketitle

\begin{abstract}
The future 21\,cm intensity mapping observations constitute a promising way to trace the matter distribution of the Universe and probe cosmology. 
Here we assess its capability for cosmological constraints using as a case study the BINGO radio telescope, that will survey the Universe at low redshifts ($0.13 < z < 0.45$). 
We use neural networks (NNs) to map summary statistics, namely, the angular power spectrum (APS) and the Minkowski functionals (MFs), calculated from simulations into cosmological parameters. 
Our simulations span a wide grid of cosmologies, sampled under the $\Lambda$CDM scenario, \{$\Omega_c, h$\}, and under an extension assuming the Chevallier-Polarski-Linder (CPL) parameterization, \{$\Omega_c, h, w_0, w_a$\}. 
In general, NNs trained over APS outperform those using MFs, while their combination provides 27\% (5\%) tighter error ellipse in the $\Omega_c-h$ plane under the $\Lambda$CDM scenario (CPL parameterization) compared to the individual use of the APS.  
Their combination allows predicting $\Omega_c$ and $h$ with 4.9\% and 1.6\% fractional errors, respectively, which increases to 6.4\% and 3.7\% under CPL parameterization. 
Although we find large bias on $w_a$ estimates, we still predict $w_0$ with 24.3\% error. 
We also confirm our results to be robust to foreground contamination, besides finding the instrumental noise to cause the greater impact on the predictions. 
Still, our results illustrate the capability of future low redshift 21\,cm observations in providing competitive cosmological constraints using NNs, showing the ease of combining different summary statistics. 
\end{abstract}

\begin{keywords}
large-scale structure of Universe -- cosmological parameters -- methods: statistical
\end{keywords}



\section{Introduction}

The production of catalogues of galaxies and clusters of galaxies through large redshift surveys is of fundamental importance to cosmology, in particular for robust measurements of the large scale density fluctuations of the Universe. 
In fact, the galaxy distribution is mainly driven by the dark matter distribution, 
and their evolution since the initial gravitational collapse is influenced by the amount of dark matter and dark energy in the Universe, making the measurements of the clustering of galaxies one of the main observational probes in cosmology. 

As an alternative to the detection of individual galaxies, the low resolution measurements of the redshifted 21\,cm line emission of the neutral hydrogen (\hi), using intensity mapping \citep[IM; see, e.g.,][]{Peterson:2006,Peterson:2009}, configures a new observational technique for tracing the large scale structure of the Universe and constraining cosmological parameters \citep{Pritchard:2012}. 
Most of the {\hi} in the late Universe is located inside the galaxies, and, as opposed to using the 21\,cm radio emission to resolve the galaxies, the IM allows measuring its overall brightness temperature fluctuations, similar to cosmic microwave background (CMB) observations, but as a function of the redshift. 
The low resolution of such technique makes it relatively cheap and allow a quick survey of large volumes of the Universe \citep{2013/battye}. 
Aiming to explore this new observable, several radio instruments, in operation and under construction, are expected to map the large scale Universe using the {\hi} IM. 
Among them are the Square Kilometre Array\footnote{\url{https://www.skatelescope.org/}} \citep[SKA;][]{2020/ska}, the Canadian Hydrogen Intensity Mapping Experiment\footnote{\url{https://chime-experiment.ca/}} \citep[CHIME;][]{2014/chime}, Five-Hundred-Meter Aperture Spherical Radio Telescope \cite[FAST;][]{2011/fast}, and the Baryon Acoustic Oscillations from Integrated Neutral Gas Observations\footnote{\url{https://www.bingotelescope.org/en/}}  \citep[BINGO;][]{2013/battye, 2015/bigot-sazy, 2021/abdalla_BINGO-project}. 
In this paper, exploring the BINGO telescope as a case study, we assess the constraining power of future {\hi} IM measurements at low redshifts ($0.13 < z < 0.45$), evaluating the performance of the joint use of summary statistics and neural networks for this task.

Inspired by the human brain, made up of connected networks of neurons, a neural network (NN) processes the information through a set of algorithms so that it can learn from examples and observation, mimicking the human learning process. 
These neurons are distributed into layers; a NN is classified as a deep learning algorithm according to the number of layers (or how dense it is), usually more than three \citep[for a review on deep learning, see][]{2015/schmidhuber}. 
A sub-field of machine learning, the NNs are pattern recognition algorithms and have been widely employed for cosmological analyses over the last years following diverse approaches. 
Commonly employed are the convolutional neural networks (CNNs), able to extract information directly from the cosmological field, e.g., as done by \citet{2018/gupta, 2019/fluri, 2019/ribli, 2020/matilla, 2022/lu} using weak lensing maps, and by \citet{2021/lazanu, 2022/villaescusa} using density fields. 
The NNs are also commonly used taking summary statistics as input data, as done, e.g., by \citet{2014/novaes, 2015/novaes} using Minkowski functionals from CMB maps, by \citet{2020/jennings} using the three-point correlation function from 21\,cm signal distribution, and by \citet{2022/perez} using three different estimators, the two-point correlation function, the count-in-cells, and void probability function; a hybrid approach combining this simple NN and a CNN is employed by \citet{2020/ntampaka} using the power spectrum. 
In addition, deep learning has been employed for data compression, aiming to extract optimal summary statistics, showing to be an efficient way for cosmological parameters inference \citep{2018/alsing,2019/alsing,2021/jeffrey}. 
We note that these are only few examples, among several other machine learning and deep learning applications for cosmological analyses \citep[for an example of a recent usage of machine learning algorithms, see][]{2022/vonMarttens}.  

Moreover, NNs can be an alternative to likelihood based analyses, performing cosmological parameter inference directly from simulations. 
The motivation for such approach is the difficulty in theoretically modelling physical aspects, such as the signature of non-linear evolution of the structures on small scales, as well as instrumental characteristics, as noise and systematic errors, which are more easily reproduced by simulations. 
Another advantage of such approach is that one can avoid making assumptions, commonly required and sometimes not completely correct, for an analytical expression for the likelihood, such as Gaussianity \citep{2021/jeffrey}. 
In fact, the highly non-Gaussian distribution of the structures, a result of the non-linear evolution of the Universe, requires the usage of higher order statistics, which, in general, has no analytical expression and, consequently, no likelihood function. 
On this context, parameter inference from simulations, or their summary statistics, through NNs algorithms, without an explicit likelihood, seems to be very advantageous for cosmological constraints.

Indeed, machine learning techniques, or in particular NNs, are very versatile tools and can be employed in uncountable ways and purposes.   Among the several possibilities of usage, we chose to explore how much cosmological information one can extract from using a simple fully connected NN trained over two summary statistics. 
Our summary statistics consist of the angular power spectrum (APS), commonly used to extract cosmological information from different data sets, and the Minkowski functionals \citep[MFs;][]{1903/minkowski, 1999/novikov, 2003/komatsu}, sensitive to higher order correlations \citep[for an example the MFs used as input to NNs, see ][]{2014/novaes, 2015/novaes}. 
The MFs are widely used to explore statistical properties of the two-dimensional CMB temperature field \citep{2003/komatsu, 2013/modest, 2019/planckVII, 2015/novaes, 2016/novaes} and the two- and three-dimensional distribution of galaxies in the Universe \citep{2007/saar,2010/kerscher,2013/choi, 2018/novaes}. 
Beyond the cosmological information contained in the power spectrum, non-gaussian information extracted through MFs can provide additional tools to differentiate between cosmological models, constraining cosmological parameters \citep{2014/shirasaki, 2015/petri} and probing modifications of the gravity \citep{2017/fang,2017/shirasaki}.
Differently from APS and bispectrum calculations, the MFs have the advantage of the additive property \citep{1999/novikov}, allowing them to be efficiently applied to small regions of the sphere, particularly useful for partial sky surveys. 
Given that the MFs have no analytical expression, modelling them and their correlation with the APS efficiently is not an easy task, in particular at low redshifts, where matter distribution is highly non-linear. 
Then, the complexity in constructing a likelihood of the data also motivates our usage of NNs in constraining cosmological parameters from APS and MFs. 

In this work, we generate a large set of lognormal simulations of the 21\,cm signal spanning a wide grid of cosmologies under two scenarios, namely, the $\Lambda$ Cold Dark Matter ($\Lambda$CDM) and an extension including dark energy (DE) equation of state (EoS) parameters. 
We also account for foreground contamination and instrumental effects, such as noise and beam size, considering the specifications for the BINGO telescope. 
We train the NN algorithm over the APS and MFs calculated from these simulations, in such a way to map these summary statistics in terms of the input cosmological parameters. 
Using each summary statistic individually and their combination, we evaluate the performance of our method in constraining cosmological parameters from each model and how the contaminant signals affect these results. 
We also investigate the sensitivity of each summary statistic with the sky coverage, evaluating our results over a larger field of view, coverage similar to what the SKA instrument will survey. 

This paper is organised as follows: 
In Section \ref{sec:simulations} we describe the 21\,cm mocks and the parameter space over which they are generated and how the simulations are constructed, accounting for the instrumental noise, the foreground components, and the foreground cleaning process. 
The summary statistics and the procedure used to obtain an optimised NN algorithm to each case investigated here, as well as the metrics used to evaluate the accuracy of our predictions, are presented in Section \ref{sec:methods}. 
We discuss our results in Section \ref{sec:results} and summarise the main conclusions in Section \ref{sec:conclusions}.

\section{Simulations} \label{sec:simulations}

In this section, we describe the simulated sky maps used in our analyses to produce the training and test data sets. 
All the simulations are produced in the {\tt HEALPix} pixelization scheme \citep{2005/gorski} with $N_{\rm side} = 256$. 
Jointly to the cosmological 21\,cm signal, we account for the expected foreground contamination, thermal noise, and sky coverage, produced according to the prescriptions provided by \cite{2021/fornazier_BINGO-component_separation, 2022/mericia_BINGO-component_separation_II, 2021/abdalla_BINGO-optical_design}, which we briefly describe below. 
We note that these effects are accounted in our simulations following the procedure described in \cite{2022/novaes-bingo}, to which we refer the reader for detailed information. 

\subsection{Cosmological signal} \label{sec:cosmo_signal}

We use the publicly available Full-sky Lognormal Astro-fields Simulation Kit \cite[{\tt FLASK};][]{2016/xavier} code to generate all the 21\,cm IM simulations employed here. 
As input to the {\tt FLASK} code we used the angular auto- and cross-power spectrum $C^{ij}_\ell$, for $i$ and $j$ redshift bins (z-bins), calculated using the Unified Cosmological Library for $C_\ell$'s ({\tt UCLCL}) code \citep{2019/loureiro, 2017/mcleod}, taking into account the redshift space distortion effect. 
A brief discussion about the theoretical $C^{ij}_\ell$ can be found in \cite{2022/novaes-bingo}, while a detailed description on how they are calculated appears in \cite{2019/loureiro}. 

The input $C^{ij}_\ell$ are calculated for a grid of cosmological parameters under two scenarios:
\begin{enumerate}
    \item[{\bf $(i)$}] the standard flat $\Lambda$CDM model, varying the parameters \{$\Omega_c, h, \Omega_b, n_s, A_s$\}, and 
    \item[{\bf $\,(ii)$}] the extension given by the Chevallier-Polarski-Linder (CPL) parameterization \citep{2001/chevallier, 2003/linder}, $w_{\rm cpl}(z) = w_0 + w_a(z/(1+z))$, where the $\Lambda$CDM model is recovered for $w_0 = -1$ and $w_a = 0$, for which we vary two more parameters, \{$\Omega_c, h, \Omega_b, n_s, A_s, w_0, w_a$\}.
\end{enumerate}
In each of the cases, we calculate the $C^{ij}_\ell$ for 800 different combinations of cosmological parameters.
For an efficient sampling of the parameters space, we employ the Latin Hypercube approach \citep{1979/mckay_LH}. 

\begin{figure}
	\includegraphics[width=\columnwidth]{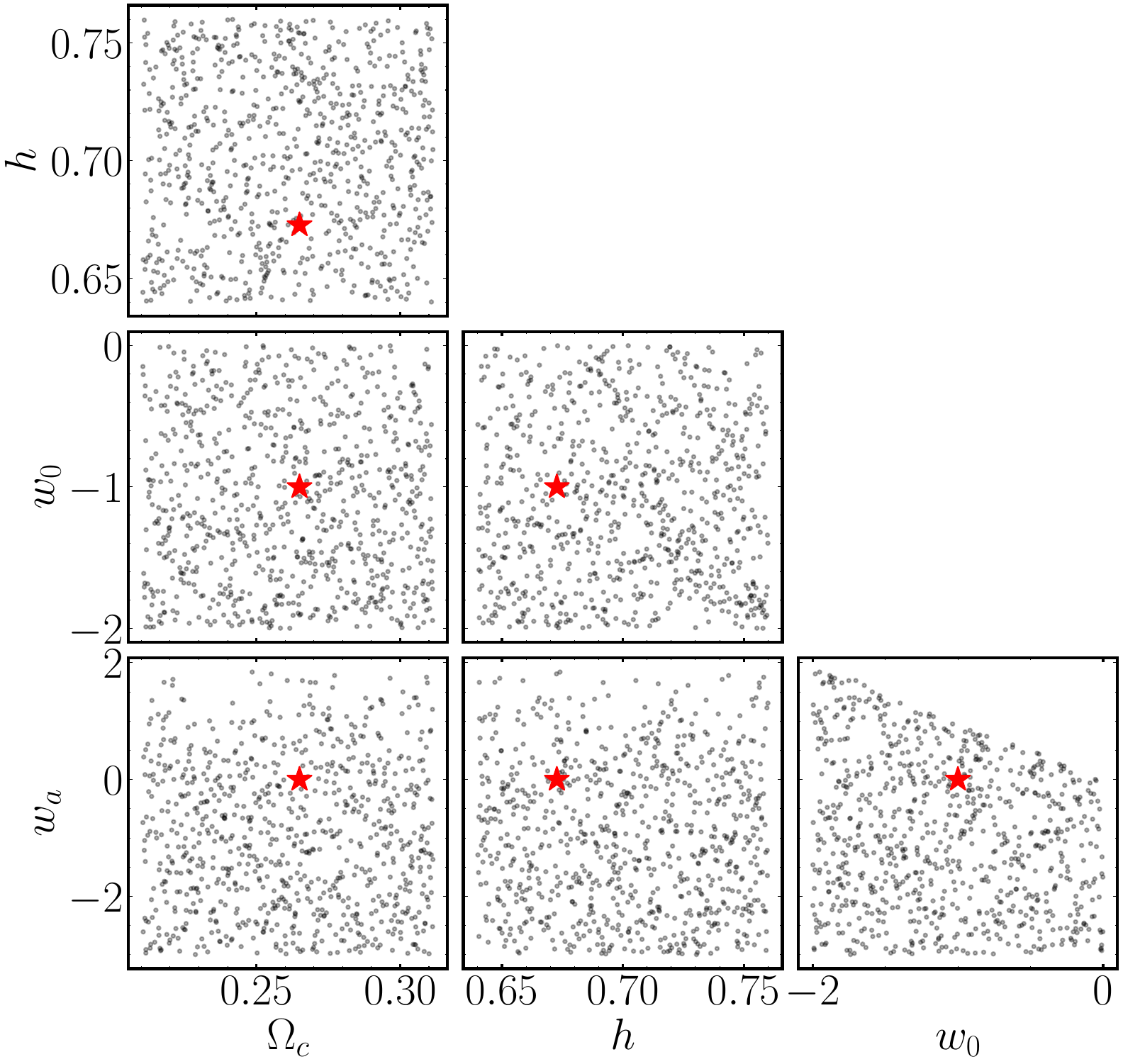}
    \caption{Latin Hypercube sampling of the parameter space (Table \ref{tab:parameters_range}). Each panel shows a grid of the 800 simulated cosmologies as grey dots. The red star show the fiducial cosmology, \{$\Omega_c, h, w_0, w_a$\} = \{$0.265, 0.6727, -1, 0$\}.  }
    \label{fig:LHsample}
\end{figure}

Although we vary all the parameters in each case, we keep $\Omega_b, n_s$, and $A_s$ values within the 1$\sigma$ error bars given by Planck 2018 \citep{2020/aghanim-planck}, while the other parameters are sampled in a broader range of values. 
This choice is because we are interested in constraining only \{$\Omega_c, h$\} and \{$\Omega_c, h, w_0, w_a$\} parameters in cases {\it (i)} and {\it (ii)}, respectively. We consider the interval of values for these parameters and their fiducial values as given in Table \ref{tab:parameters_range}; the parameter space of interest is also shown in Figure \ref{fig:LHsample} (hereafter, a given combination of parameters, i.e., a point in the grid, is simply refereed as a cosmology). 
For each cosmology we calculate $C^{ij}_\ell$ and from them generate 12 realisations of 21\,cm mocks. 
Here, one mock refers to 30 tomographic maps, one for each z-bin in the total range of $0.127 < z < 0.449$. 

It is worth to mention that the {\tt FLASK} code, by construction, follows predefined statistical properties, such as the power spectrum and multivariate log-normal model, as stated by \citet{2016/xavier}. 
That is to say that, although it has the great advantage of allowing the production of a large set of mocks in a short time, these simulations may not reproduce realistically the three-point information from the 21\,cm cosmological signal. 
In this sense, the {\tt FLASK} mocks are likely enough to evaluate our methodology using the APS, but may prevent us of exploring the full potential of the MFs, as will be pointed out in Section \ref{sec:results}.

\begin{table}
    \linespread{1.2}
    \selectfont
	\centering
	\caption{Interval for sampling the cosmological parameters. Last column show their fiducial values \citep{2018/planck-VI-CosmoParams}. }
	\label{tab:parameters_range}
	\begin{tabular}{lcc} 
		\hline
		Parameter & Interval & Fiducial \\
		\hline
		$\Omega_b$ & [0.0487, 0.0493] & 0.049\\
		$\Omega_c$ & [0.21, 0.31]     & 0.265 \\
		$h$        & [0.64, 0.76]     & 0.6727 \\
		$n_s$      & [0.9605, 0.9693] & 0.9649 \\
		$\ln(10^{10} A_s)$ & [3.029, 3.061] & 3.045\\
		$w_0$      & [-2.0, 0.0]      & -1 \\
		$w_a$      & [-3.0, 2.5]      & 0 \\
		\hline
	\end{tabular}
\end{table}

\subsection{Observational characteristics} \label{sec:observations}

\paragraph*{($i$) Foreground contamination:} 
As in \cite{2022/mericia_BINGO-component_separation_II, 2021/fornazier_BINGO-component_separation}, the foreground components contributing in the BINGO frequency range, $980 - 1260$ MHz, are generated by the Planck Sky Model software \citep[PSM;][]{2013/delabrouille}. 
These include the Galactic synchrotron\footnote{Here produced by a power law with non-uniform spectral index as given by \cite{2002/giardino} model. In Section \ref{sec:robustness_foregrounds} we consider an alternative model to evaluate the robustness of our results to foreground contamination.} and free-free emissions, the main contaminants in this frequency band, as well as, the thermal dust and anomalous microwave emission, beside the extragalactic contribution from thermal and kinetic Sunyaev-Zel'dovich effects, and unresolved point sources. 
We refer the reader to \citet[][Section 4.1.2]{2022/novaes-bingo} and \cite{2022/mericia_BINGO-component_separation_II}, as well as references therein, for information on the specific configuration of the PSM code for simulating each foreground component. 

\paragraph*{($ii$) Sky coverage:}
BINGO will survey a sky area of 5324 deg$^2$, corresponding to the sky fraction of $f_{\it sky} \sim 0.14$, covering a 14.75$^\circ$ stripe centred at the declination $\delta = -15^\circ$ . 
The specific footprint is reproduced in our simulations using an appropriate mask of the sky coverage \citep{2022/mericia_BINGO-component_separation_II}. 
This mask is updated to cut out a region with strong Galactic foreground emission. 
Using the {\tt NaMaster} code\footnote{\url{https://namaster.readthedocs.io/en/latest/}} \citep[][see also Section \ref{sec:summary_stats}]{2019/alonso-namaster}, the mask is apodized with a cosine transition of 5 deg, in order to avoid the impact of sharp edges. 
The final mask, after apodization and Galactic cut, has an effective sky fraction of $f_{\it sky} \sim 0.09$.

\paragraph*{($ii$) Instrument:} 
In its Phase 1, BINGO is expected to operate with 28 horns and a system temperature of 70 K \citep{2021/wuensche_BINGO-instrument}. 
With 5 years of observation, the horns positions are slightly shifted each year in order to perform a more homogeneous scan of the innermost BINGO area \citep{2021/abdalla_BINGO-optical_design}. 
As detailed in \citet{2021/fornazier_BINGO-component_separation}, such specifications allow estimating the amplitude (root-mean-square - RMS) by pixel of the BINGO thermal (white) noise, the same for all the frequency bins. 
Then, multiplying the RMS map by Gaussian distributions of zero mean and unitary variance, we generate as many realisations of noise maps as the number of simulations employed here. 
Before adding the thermal noise, we account for the instrumental beam, assumed to be approximately Gaussian with same full width half maximum $\theta_{\rm FWHM} = 40$ arcmin for all frequency bins.

\subsection{Foreground cleaning process} \label{sec:foreg_clean_proc}

An essential procedure before using the 21\,cm observations for cosmological analyses is the removal of the foreground contamination, with amplitude of $\sim 10^4$ larger than the 21\,cm signal in BINGO frequencies (and even larger at lower frequencies, such as those of SKA). 
For this, we employ the generalized needlet internal linear combination ({\tt GNILC}) method \citep{2011/remazeilles}, a non-parametric component separation technique, which has been shown to be efficient when applied to 21\,cm IM simulations \citep{2016/olivari,2021/liccardo_BINGO-sky-simulation, 2021/fornazier_BINGO-component_separation, 2022/mericia_BINGO-component_separation_II}. 
We note that {\tt GNILC} recovers the cosmological signal plus noise maps in each frequency. 
For detailed explanation about {\tt GNILC}, we refer the reader to \cite{2011/remazeilles, 2016/olivari}. 
Technical details about this foreground cleaning procedure implemented on BINGO simulations, similar to those employed here, are provided by \cite{2021/liccardo_BINGO-sky-simulation, 2021/fornazier_BINGO-component_separation, 2022/mericia_BINGO-component_separation_II}.

As in \cite{2022/novaes-bingo}, the large amount of simulations required here makes unfeasible the application of {\tt GNILC} over each of them. 
For this reason, we estimate the residual foreground signal expected to remain after the component separation process and add it to our simulations. 
We apply {\tt GNILC} to ten complete simulations, including all observational characteristics as discussed in the previous subsections, and from each of them we estimate the foreground residual maps. 
This way, our final simulations are constructed by adding the realistic foreground residual contribution to each of the 21\,cm mocks (with the beam effect already accounted for), along with the thermal noise. 
The foreground residual maps are repeated every ten mock realisations. 
Unless stated otherwise, our analyses are applied to these simulations.

\section{Methods} \label{sec:methods}

\subsection{Summary statistics} \label{sec:summary_stats}

\paragraph*{Angular power spectrum}
Recognised as a powerful statistic commonly used to constrain cosmology, the APS is calculated as the average of the $a_{\ell m}$ coefficients of the decomposition of a temperature fluctuation (or projected density fluctuation) field into spherical harmonics $\hat{C}_{\ell} = \langle a_{\ell m} a_{ \ell m}^* \rangle$. 
Here we employ the pseudo-$C_\ell$ method as implemented in the {\tt NaMaster} code \citep{2019/alonso-namaster} to estimate the APS. 
This formalism relates the observed APS, $\hat{C}_\ell$, to the true spectrum $C_\ell$ as
\begin{equation} \label{eq:namaster}
    \hat{C}_{\ell} = \sum_{\ell'} \mathcal{M}_{\ell \ell'} C_{\ell'},
\end{equation}
where the mode-coupling matrix $\mathcal{M}_{\ell \ell'}$ is determined by the mask geometry \citep{2002/hivon, 2005/brown}. 
This matrix is analytically calculated, and the APS is estimated by inverting Equation \ref{eq:namaster}. 
Given the sky fraction assumed here ($f_{\rm sky} = 0.13$), we calculate the $C_\ell$ at bins with width $\Delta \ell = 10$ \citep{2022/novaes-bingo}, which makes the mode-coupling matrix invertible. 
It means that we use a total of 76 data points (multipole bands) linearly spaced in $\ell$, with $2 \leq \ell \leq 761$ (with maximum multipole as given by the pixelization of the simulations, $N_{side} = 256$, and the multipole binning, $\Delta \ell = 10$). 
Here we consider only the auto-APS; the constraining power of the cross-APS between different z-bins will be assessed in future work.

\paragraph*{Minkowski functionals}
Unlike the APS, besides informing about spatial correlation of a random field, the MFs can also provide morphological information and map the shape of structures.
The morphological properties of a given field in a $N$-dimensional space can be completely described using $N+1$ MFs \citep{1903/minkowski}. 
Then, a two-dimensional 21\,cm temperature fluctuations field, $\delta T$, with variance $\sigma_0^2$, would be completely described by three MFs, namely, the Area ($V_0$), Perimeter ($V_1$), and Genus ($V_2$), given by \citep{1999/novikov, 2003/komatsu, 2013/ducout, 2018/novaes}

\begin{align}
\! V_0(\nu) &= \frac{1}{4 \pi} \int_{\Sigma} d\Omega \, ,  \label{eq:v0}\\
\! V_1(\nu) &= \frac{1}{4 \pi} \frac{1}{4} \int_{\partial\Sigma} dl \, ,  \label{eq:v1}\\
\! V_2(\nu) &= \frac{1}{4 \pi} \frac{1}{2 \pi} \int_{\partial\Sigma} \kappa~dl  \label{eq:v2}\, ,
\end{align}
where $d \Omega$ and $dl$ are the elements of solid angle and line, respectively, and $\kappa$ is the geodesic curvature \cite[for details see, e.g.,][]{2013/ducout}.
These MFs are calculated for an excursion set defined as $\Sigma \equiv \{ \delta T > \nu \sigma_0 \}$, i.e., the set of connected pixels exceeding a given $\nu$ threshold, whose boundary is $\partial \Sigma \equiv \{ \delta T = \nu \sigma_0 \}$. 
The first two MFs measure the area and the contour length of the excursion set, and the last one gives the difference between connected areas above the threshold $\nu$ and below it in the excursion. 

It is worth to mention that analytical expressions for the MFs are known only for Gaussian and weakly non-Gaussian cases.
They can be written as $V_k = A_k {\rm v}_k$, where, for Gaussian fields, ${\rm v}_k = {\rm v}_k^G = H_{k-1}(\nu)$; $H_{k}(\nu)$ is the $k$-th Hermite polynomial. 
The amplitude $A_k$ depends on the shape of the APS. 
For non-Gaussian fields, ${\rm v}_k$ can be expanded in a Taylor series, ${\rm v}_k = {\rm v}_k^G + {\rm v}_k^{NG}$, where the non-Gaussian correction terms, represented by ${\rm v}_k^{NG}$, depend on the higher order moments of the field \citep{2015/petri, 2013/ducout, 2010/matsubara}. 
Although the first order corrections to the MFs can be analytically obtained, the perturbative solution is not enough for large non-Gaussianity, as is the case of the 21\,cm signal at low redshifts, and the series does not converge.

Here we numerically calculate the MFs using the code provided by \cite{2013/ducout} and \cite{2012/gay}.
The number of $\nu$ thresholds is fixed at 31, defined dividing the range $[\nu_{min},\nu_{max}] = [-2.5,6.0]$ in equal parts.

\begin{figure}
 	\includegraphics[width=1\columnwidth]{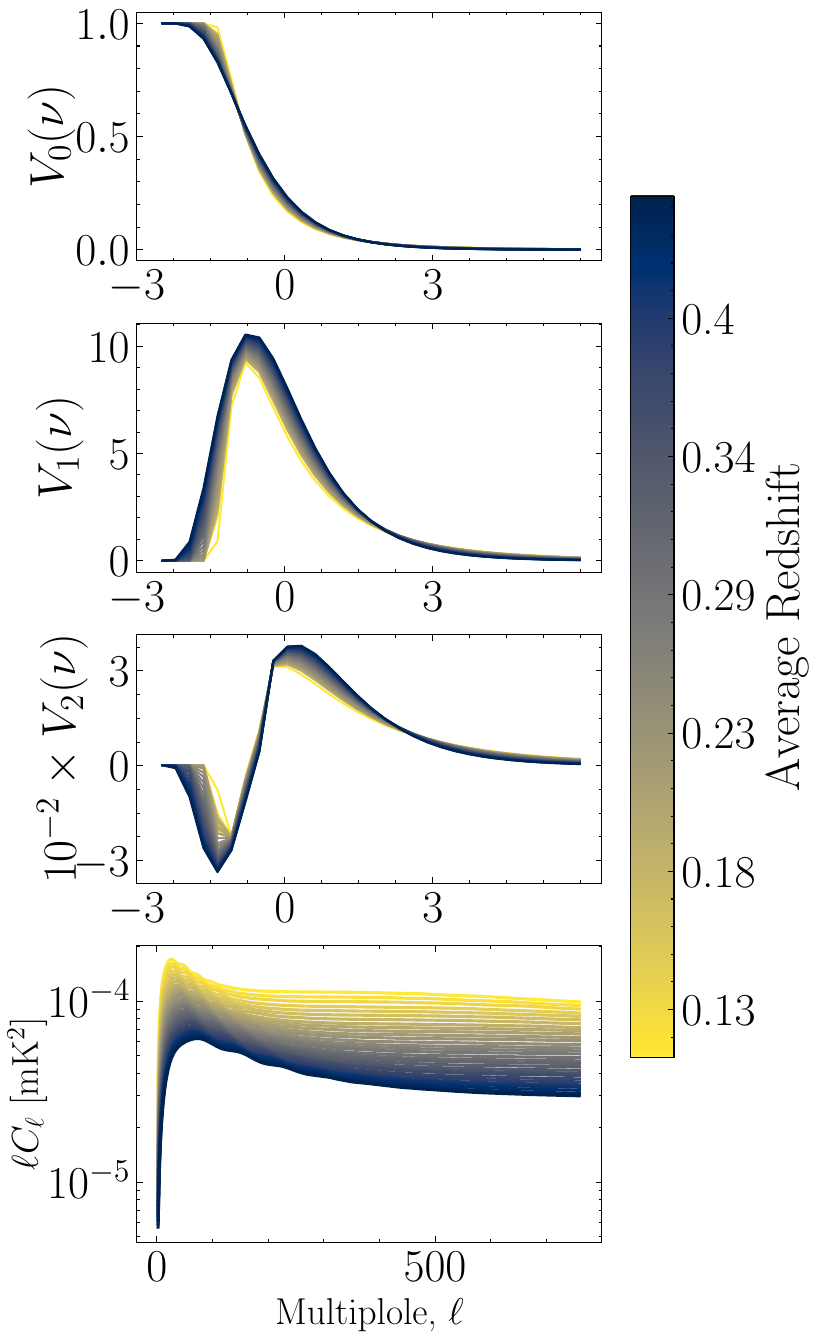}\\
    \caption{Average MFs (the three upper panels, from top to bottom, are the Area, Perimeter and Genus) from 1000 clean 21\,cm mocks, accounting for the BINGO sky fraction and beam, and theoretical APS (bottom panel), for the fiducial cosmology (Table \ref{tab:parameters_range}). The lines correspond to the 30 tomographic bins, coloured according to the average redshift of each z-bin. 
    }
    \label{fig:ClsMFs-curves}
\end{figure}

\vspace{0.3cm}
As illustration, Figure \ref{fig:ClsMFs-curves} shows, for the fiducial cosmology, the average MFs from 1000 realizations of clean 21\,cm simulations, as well as the theoretical APS, showing their dependence with the redshift. 

\subsection{Neural network}

Working as a regression technique, a NN is composed by processing units, the neurons, organised in layers: one input layer, fed by the input (features) and target information, one or more hidden layers, and the output layers. 
In a fully connected NN, each neuron in a layer is connected to those in the next layer, allowing the information to be processed and to propagate until the output layer. 
Following a supervised learning, the outputs of this process are compared to the target values (true information) estimating an error (or loss) function to measure the performance. 
The errors are propagated to the input layer to adjust the parameters of the NN algorithm so that the error function is minimised, performing an iterative process until the error achieve a given threshold, following a backpropagation training process.
A detailed description about the NNs is provided by, e.g., \cite{2020/choudhury,2020/jennings}.

Next subsections describe how the summary statistics are organised to feed the NN, as well as the architecture and how the hyperparameters (number of layers, number of neurons, among others) are chosen. 
We also present the accuracy metrics used to evaluate the performance of the NNs in constraining the cosmological parameters.

\subsubsection{Training and test data sets} 

For each of the scenarios we investigate here, cases {\it (i)} \{$\Omega_c, h$\}  and  {\it (ii)} \{$\Omega_c, h, w_0, w_a$\}, we generate 12 realisations of 21\,cm mocks for each of the 800 cosmologies. 
We find that a set of $n = 12$ maps for each cosmology seems to be enough for the NN to account for the cosmic variance (CV; tests have shown that for $n \geq 4$ the accuracy of the NN plateaus and oscillates around an average value). 
We calculate the APS and the MFs statistics for each of the simulations, at each z-bin. 
The summary statistics are the features feeding our NNs algorithms and the corresponding cosmological parameters values are the targets, or outputs, of the NNs, i.e., the quantities we want to predict using the trained NN. 
These data are split so that 64\%, 16\% and 20\% of the 800 cosmologies are used for training, validation and testing of the NN, respectively. 
It means that APS and the MFs calculated over mocks from a given cosmology used for training or validation are not employed for testing the performance of the NN. 
To guarantee that our results do not depend on a particular (random) choice of train and validation sets, we use the cross-validation procedure. 
For this, we split the train+validation data set into 5 smaller data sets, training the NN in 4 of them and validating on the remaining set. 
These training and validation steps are repeated 5 times, each of them excluding one of the 5 smaller sets, and the performance value is given by the average of the values measured at each of the 5 training processes. 
We note that, before feeding the NN, the features are re-scaled in such a way that all the values lie in the range [0,1], a procedure commonly employed to improve the efficiency during the training process \citep{2020/jennings}. 

The training+validation set is defined as $T\{X,y\} = T\{X_j,y_j\}$, allowing to map the features, i.e., the summary statistics, $X$, in terms of the targets, given by the corresponding cosmological parameters, $y$. 
For the $j$th realisation the features appear as
\begin{equation}
\begin{split} \label{eq:features}
      X_j &= \{ (C_\ell^i), (V^i_k) \}|_{j}, \\
      &= \{ (C_\ell^1, C_\ell^2, ..., C_\ell^m), (V^1_k, V^2_k, ...V^m_k) \}|_{j},
\end{split}
\end{equation}
where $V_k(\nu) \equiv (V_0(\nu), V_1(\nu), V_2(\nu))$, representing the Area, Perimeter, and Genus estimators, respectively, and the index $i= 1, 2, ..., m$ runs over the $m=30$ z-bins. 
This means that, for each realisation, $X_j$ is a vector with $m \times [\mbox{76 multipole bands + 3 MFs} \times 31 ~\nu ~\mbox{thresholds}]$ elements. 
Hereafter, we refer to the three MFs combined as $V_k$.
The features are associated to the targets,
\begin{equation}
    y_j = \{\theta_p\}|_j,
\end{equation}
where, for case \textit{(i)}, $p = 0, 1$ with $\{\theta_p\} = \{\Omega_c, h\}$ and, for case \textit{(ii)}, $p = 0$ with the target given by one of the parameters at a time\footnote{When a parameter is not well determined, as is the case, in particular, for $w_a$ (see Section \ref{sec:results}), the improvement from the other parameter has no impact on the loss function. We find more accurate predictions for case \textit{(ii)} when training NN algorithms individually for each cosmological parameter. }, $\{\theta_p\} = \{\Omega_c\}, \{h\}, \{w_0\}, \mbox{ or } \{w_a\}$, i.e., the NN algorithm is trained over individual cosmological parameters. 
We test the efficiency of each estimator, $C_\ell$ and $V_k$, individually, $X = \{ (C_\ell^i) \}$ and $X = \{ (V^i_k) \}$, as well as their combination, $X = \{ (C_\ell^i), (V^i_k) \}$, changing equation \ref{eq:features} accordingly.

Our main analysis considers a compression of all the 30 z-bins into 5 redshift ranges, given by a simple average of the summary statistics over every 6 consecutive z-bins (the motivation for this choice is discussed in Section \ref{sec:sensitivity_zbins}). 
Therefore, the $i$ index runs over $m = 5$ redshift intervals, or subsets of the z-bins.

\subsubsection{Architecture} 

For each of the tests reported here, we define a particular architecture by searching for an optimised set of hyperparameters. 
For this we use the {\tt Optuna} package \citep{2019/akiba_optuna}, able to automatically search the values of the hyperparameters, from a predefined search space, by minimising/maximising a given objective function. 
Here we use a loss function, chosen to be the commonly used mean square error (MSE), given by
\begin{equation} \label{eq:mse}
    \mathcal{L} = \frac{1}{N} \sum_{j=1}^{N} ( y^{\rm Pred} - y^{\rm True} )^2,
\end{equation}
for a total of $N$ simulations in the validation data set, with the `True' and `Pred' superscripts referring to the real (input) values of the cosmological parameters, used for generating the simulations, and the predicted values by the NN, respectively. 
Then, at each trial, i.e., after training the NN with a given set of values for the hyperparameters, a validation loss (a score) $\mathcal{L}$ is returned, a procedure repeated in order to search for the set of hyperparameters minimising it. 
For all our tests, the number of trials is limited to 500, since cases {\it (i)} and {\it (ii)} usually does not need more than 300 and 400 trials, respectively. 

Our NN algorithms are constructed optimising the following main hyperparameters, considering the searching spaces as defined below:
\begin{itemize}
    \item number of neurons in the hidden layers: [1,500];
    \item number of layers: [1,3];
    \item activation function: {\tt ReLu}, {\tt tanh};
    \item optimiser: {\tt Adam}, {\tt SGD};
    \item learning rate: [$10^{-4}, 10^{-2}$] and [$10^{-5}, 10^{-1}$] for {\tt Adam} and {\tt SGD} optimisers, respectively;
    \item momentum (only for {\tt SGD}): [$10^{-4}, 10^{-2}$];
    \item batch size: [50, 500];
    \item number of epochs: [50, 500];
    \item weight decay: [$10^{-4}, 10^{-2}$]. 
\end{itemize}
In the case of {\tt Adam} optimiser, the beta parameters are fixed to the default values $\beta_1,\beta_2 = 0.9, 0.999$. 
The optimisation processes usually finds 1 to 4 hidden layers with no more than 300 neurons, taking nine to fifteen hours to perform the 500 trials on 56 cores of a processor Intel Xeon Gold 5120 2.20 GHz and 512 GB of RAM.

\subsubsection{Accuracy metrics} 

To evaluate the performance of the NNs constructed from the training process, we quantify how accurate are the predicted values, $y^{\rm Pred}$, with respect to the true, or input, values, $y^{\rm True}$, from the test set, $t\{X\}$. 
For this we used four different metrics to characterise the accuracy of the predictions, for each cosmological parameter, evaluating  different aspects of the results. 

The first metric is the root-mean-squared error, defined by the square root of Equation \ref{eq:mse}, RMSE~$ = \sqrt{\mathcal{L}}$, for the $N$ simulations in the test set.
It quantifies the overall accuracy (total error) of the predictions, regardless of the origin, i.e., whether it is inherent to the NN algorithm or it is introduced by the CV of the cosmological signal. 
The second metric is the standard deviation of the predicted values averaged for all the $n_c = 160$ different cosmologies, 
\begin{equation}
    \langle \sigma \rangle = \frac{1}{n_c} \sum_{c=1}^{n_c} \Bigg[ \sqrt{ \frac{1}{n} \sum_{i=1}^{n} (y^c_i - \langle y^c \rangle )^2 } \Bigg] \, ,
\end{equation}
for a total of $n$ simulations from each $c$ cosmology; this metric quantifies the uncertainty associated to the CV. 
The third metric is the slope of the straight line fitted over the data points, relating predictions and real values, $y^{\rm Pred} \times y^{\rm True}$. 
It measures the bias appearing in the predictions when the NN is not able to effectively learn the relation between the summary statistics and the cosmological parameters. 
A slope near to 1 indicates a small bias and accurate predictions, while a slope near to 0 reflects a large bias and indicates that the NN is predicting values close to the mean of the parameters ranges. 
Such large bias is a common behaviour of the NN when it is not able to efficiently map the features in terms of the targets \citep[see, e.g.,][]{2022/lu}; the NN predicts values near to the mean of the parameter range because, in doing so, it artificially reduces the loss function. 

The last metric quantifies also the overall accuracy of our results, but considering the relationship between pairs of cosmological parameters. 
This metric is the area of the 1$\sigma$ error ellipses obtained from the distribution of $\Delta y = y^{\it True} - y^{\it Pred}$, the difference between true and predicted values, in the $\Delta y_{p}-\Delta y_{p'}$ plane, where $p$ and $p'$ indicate two different cosmological parameters.

\section{Results and discussions} \label{sec:results}

Here we present the results obtained evaluating the performance of the method in constraining \{$\Omega_c, h$\} and \{$\Omega_c, h, w_0, w_a$\} sets of parameters using the respective NNs trained over our simulations (cosmological 21\,cm signal + thermal noise + foreground residual). 
We evaluate the performance of the method using different features, the summary statistic, $V_k$ and $C_\ell$, individually and their combination, $V_k$+$C_\ell$. 
We also investigate how our results are affected by each contaminant signal, thermal noise and foregrounds, and their robustness to the foreground characteristics.
For the clean 21\,cm simulations, we investigate the sensitivity of the results to the sky coverage and to specific redshift ranges. 
We employ an optimised set of hyperparameters for each of these tests. 

\subsection{Parameters prediction under the $\Lambda$CDM model}  \label{sec:2parameters}

\begin{figure}
	\includegraphics[width=0.5\columnwidth]{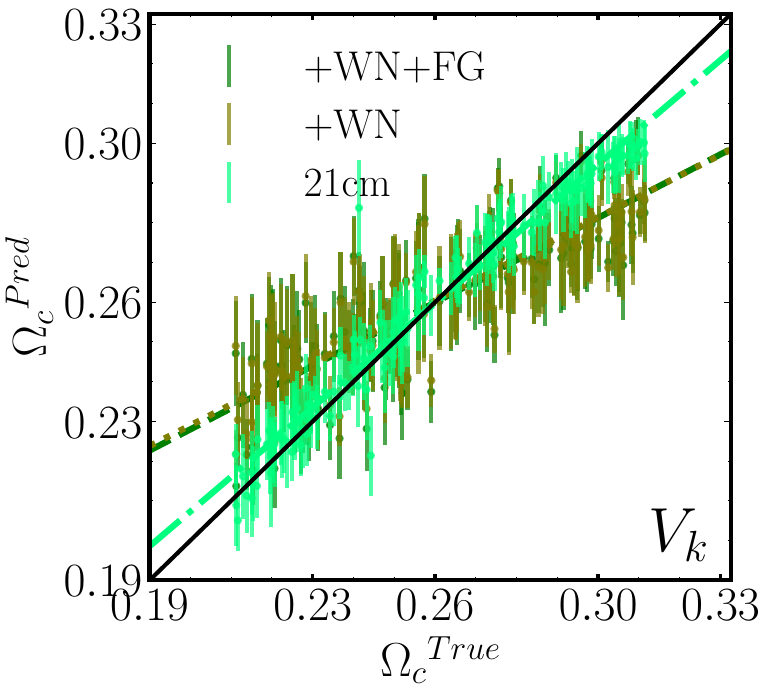}\includegraphics[width=0.5\columnwidth]{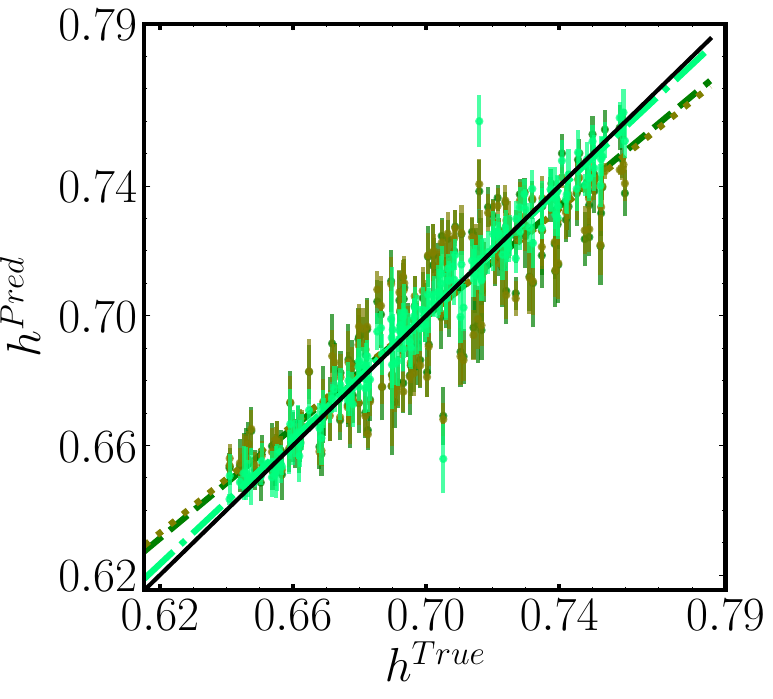}\\
	\includegraphics[width=0.5\columnwidth]{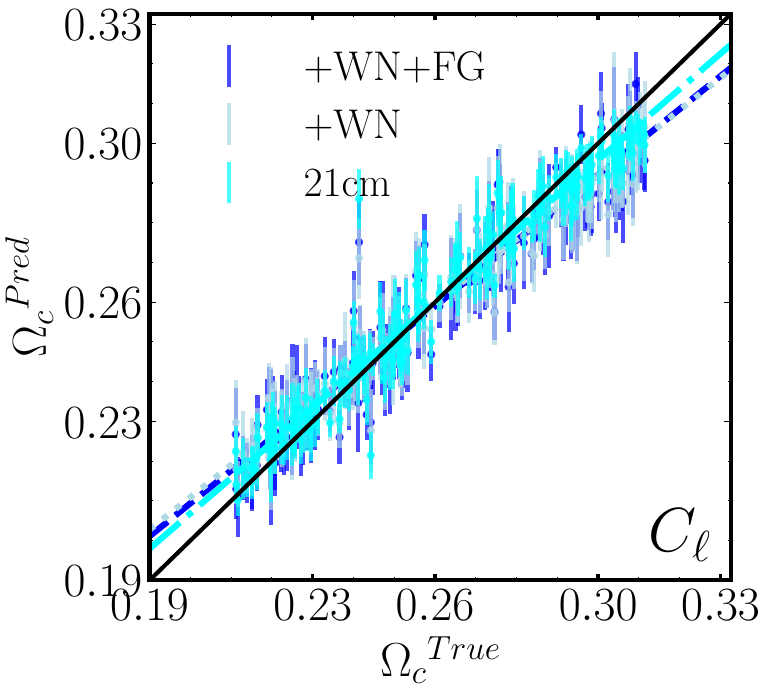}\includegraphics[width=0.5\columnwidth]{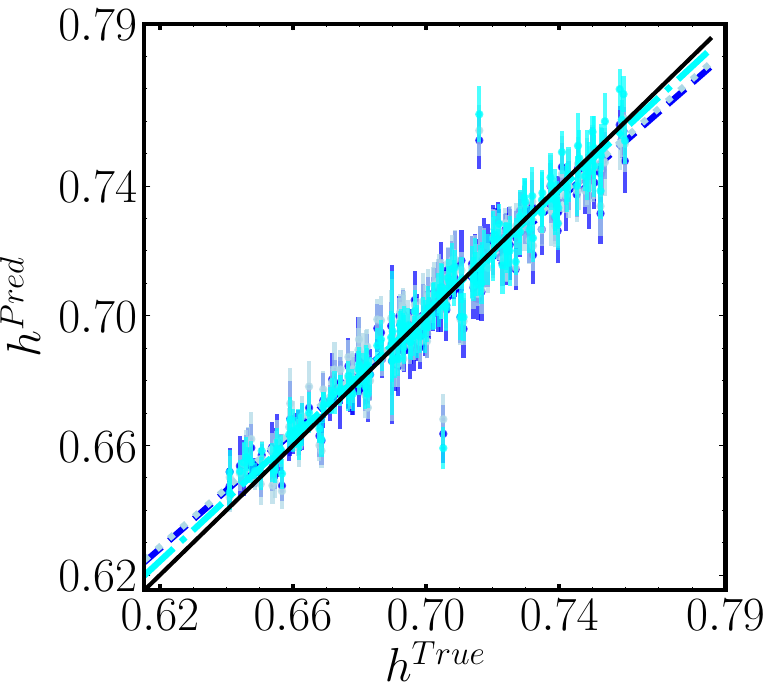}\\
	\includegraphics[width=0.5\columnwidth]{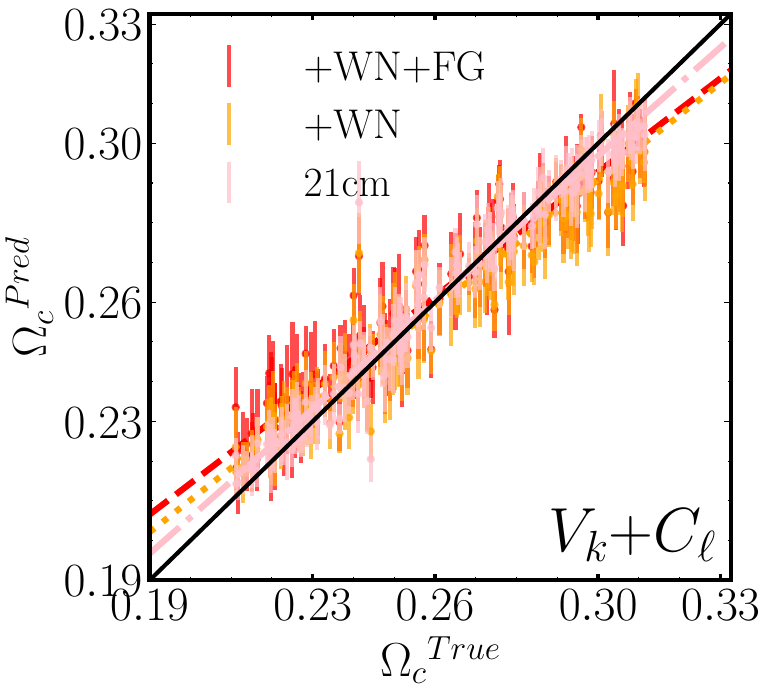}\includegraphics[width=0.5\columnwidth]{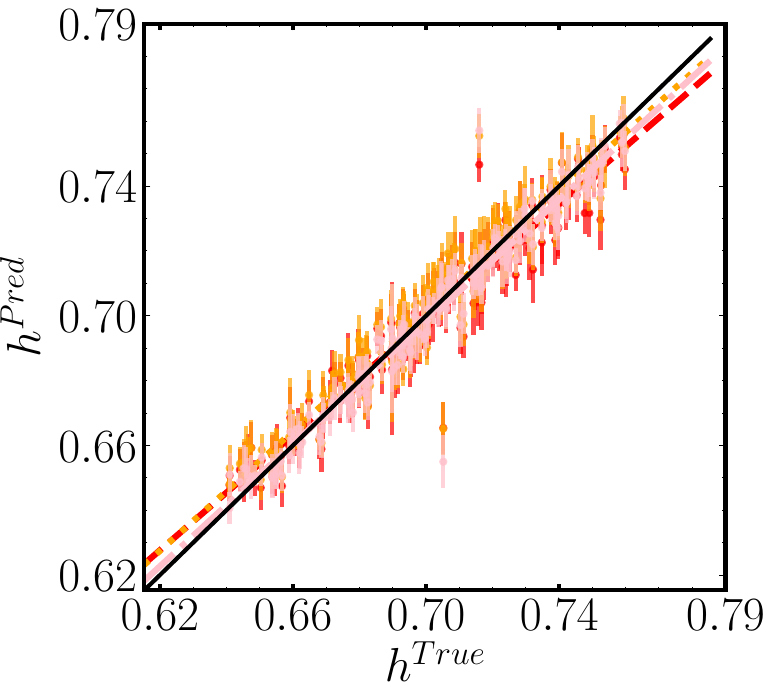}
    \caption{Predicted versus the true cosmological parameters values, showing results from case {\it (i)}, for \{$\Omega_c, h$\} parameters. Each point and error bar correspond to the average and standard deviation of the $n=12$ predictions within the same cosmology. The different colours, as labelled in the first panel of each row, represent a type of simulation, the clean 21\,cm mocks, these mocks contaminated by thermal noise (+WN), and contaminated by foreground residual, along with the noise (+WN+FG). The dot-dashed, dotted and dashed lines are the linear fitting of the corresponding coloured data points. From top to bottom, each row shows results using the MFs, the $C_\ell$, and their combination, $V_k$+$C_\ell$. }
    \label{fig:2parameters}
\end{figure}

For each of the 800 cosmologies sampling the two-parameter space, case {\it (i)} \{$\Omega_c, h$\}, we generate 12 realisations of the cosmological 21\,cm signal, then include the effect of the instrumental beam and add the contribution of thermal noise and foreground residual. 
We use the APS and MFs calculated from these simulations to train and validate the NN. 
Applying the trained NNs to the respective test data sets, we obtain the predicted values of the \{$\Omega_c, h$\} cosmological parameters. 
These predictions are compared to the true values in Figure \ref{fig:2parameters}, for each summary statistic individually and combined, showing also the best linear fit over the data points, whose slope values are summarised in Table \ref{tab:2parameters}. 
The standard deviation over the $n=12$ predicted values within each cosmology, represented by the error bars in Figure \ref{fig:2parameters}, averaged over the cosmologies from the test sets, $\langle \sigma \rangle$, are also presented in Table \ref{tab:2parameters}, as well as the RMSE values. 
The RMSE, a measurement of the overall error, is also calculated for the training data set (appearing in parentheses in the tables) so that we can assess overfitting. 
A significantly larger RMSE value from test data compared to that from training data would indicate overfitting. 
For all our analyses these values seem to be in reasonable concordance, then we consider that no overfitting occurs. 
Although omitted here, the other metrics obtained from training data sets also lead to the same conclusion.

From the third part of Table \ref{tab:2parameters}, one can see that, regardless of the features, $h$ is always better estimated than the $\Omega_c$ parameter.
For both parameters, Figure \ref{fig:2parameters} show that, given the error bars, the predictions are consistent with the true values, showing that the NNs are able to learn the relations between the summary statistics and cosmological parameters, specially when using $C_\ell$ or $V_k$+$C_\ell$.

The results from the accuracy metrics $\langle \sigma \rangle$, slope, and RMSE show that the APS outperforms the MFs in estimating both $\Omega_c$ and $h$ parameters.
Their combination, however, does not seem to provide a significant improvement on the results, showing slightly smaller (larger) RMSE values for $h$ ($\Omega_c$) parameter with respect to that from $C_\ell$, while the slope values are slightly smaller for both cosmological parameters. 
Still, for $V_k$+$C_\ell$, the RMSE for $\Omega_c$ and $h$ parameter are 0.013 and 0.011, respectively, representing 4.9\% and 1.6\% fractional errors about the fiducial values $\Omega_c = 0.265$ and $h = 0.6727$. 
As pointed by \citet{2022/perez}, for $\Omega_c$ and $h$, a RMSE value close to 0.025 and 0.030, respectively, would indicate inaccuracy of the method in predicting these parameters, since the error would amount to half of the sampling interval (Table \ref{tab:parameters_range}).
In \citet{2021/costa_BINGO-forecast}, the BINGO collaboration employs  the Fisher matrix formalism in a cosmological forecast using the APS and find a 1$\sigma$ constraint of 19\% fractional error for $h$, under the $\Lambda$CDM scenario, which is reduced to 0.9\% when adding Planck measurements. 
Although the authors explore a different set of cosmological parameters and prior ranges, so the comparison with our results is not straightforward\footnote{For examples comparing the two type of analyses and error estimates, see \cite{2022/villaescusa}.}, the significantly smaller error found here for the Hubble parameter indicates how competitive our constraints can be and suggests that the error bars can be reduced by combining different data sets to feed NN algorithms.  
Also, although one can say that our analyses combine BINGO and Planck results, given that we use Planck constraints to define the sampling interval for $\Omega_b, n_s$, and $A_s$, the parameters of interest in case {\it (i)} are sampled in a significantly broader range.

To help understand our results, in Appendix \ref{appendixA} we investigate how each of the summary statistics are modified by varying each cosmological parameter. 
Figure \ref{fig:dif_MFs}, illustrating the effect of changing the cosmological parameters amplitude in 1\%, show that all three MFs and APS seem to be more sensitive to the Hubble parameter $h$ than to the dark matter density parameter $\Omega_c$, given the amplitude of the curves. 
This explains the better accuracy metric for $h$ compared to $\Omega_c$. 
Also, the higher sensitivity of the MFs to the Hubble parameter help explaining why the combination $V_k$+$C_\ell$ improves, even if just slightly, only the $h$ parameter predictions. 
However, it remains to be better understood the reason why the combinations of the summary statistics does not provide more significant improvements with respect to their individual usage. 

In order to enlighten such finding, one should remember the dependency among the two summary statistics (see Subsection \ref{sec:summary_stats}).  
This relationship is given by an integration over several scales of the $C_\ell$ \citep{1999/novikov}, possibly losing information, in such a way that, even when applied to Gaussian fields, the two statistics may not carry the same information. 
For the case of the non-Gaussian 21\,cm field, although our simulations may not reproduce a realistic three-point information (see discussion in Subsection \ref{sec:cosmo_signal}), the MFs still describe their log-normal characteristic, unlike the APS. 
In this sense, even if the log-normal information is not enough to significantly improve the cosmological predictions from MFs with respect to APS, one cannot guarantee that the two summary statistics would have similar performances or that their combination would lead to expressive improvements.

\begin{figure}
	\includegraphics[width=\columnwidth]{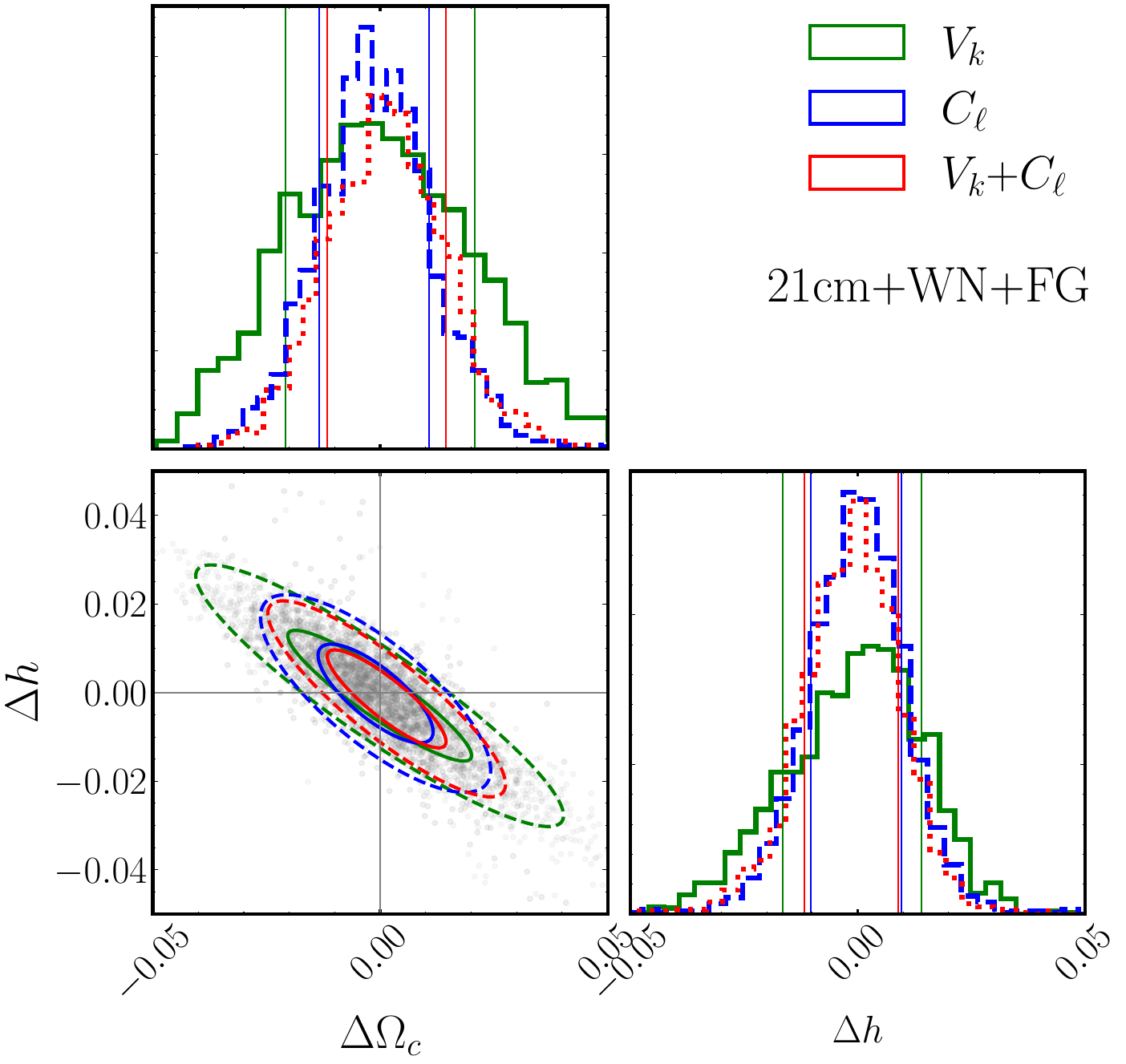}
    \caption{Error ellipses for the \{$\Omega_c, h$\} parameter space predictions, showing the distribution of the difference between true and predicted cosmological parameter, $\Delta h = h^{\rm \it True} - h^{\rm \it Pred}$ and $\Delta \Omega_c = \Omega_c^{\rm \it True} - \Omega_c^{\rm \it Pred}$. In all cases, the solid and dashed lines represent the 1$\sigma$ and 2$\sigma$ contours, respectively. The corresponding projected 1D distributions are also shown. The different colours show results from different summary statistic and their combination.}
    \label{fig:triangle_2params}
\end{figure}

In terms of the $\Delta\Omega_c$-$\Delta h$ plane (or just $\Omega_c$-$h$ plane, for simplicity), presented in Figure \ref{fig:triangle_2params}, where $\Delta$ represents the difference between true and predicted parameters, the error ellipses for the $V_k$, $C_\ell$, and $V_k$+$C_\ell$ statistics show that the last one seems to provide the smallest contour area. 
This is confirmed by the results summarised in Table \ref{tab:area_MFsAPS}, showing that the area of the error ellipse obtained using the $C_\ell$ are reduced by $\sim 27\%$ when using the combination $V_k$+$C_\ell$, while $V_k$ are again the least restrictive statistic. 
However, Table \ref{tab:area_MFsAPS} also shows that, differently from the BINGO simulations, when analysing the (unrealistic) clean 21\,cm realisations the MFs provide better constraints than the APS; their combination provides reduction of $\sim 19\%$ in the contour area with respect to MFs only. 
Such result has two main indicatives. 
First, the combination of the summary statistics seems to be even more advantageous in the presence of contaminant signals. 
Second, a possible imprecision of three-point information of our 21\,cm log-normal realisations is not the only explanation for the best performance of the $C_\ell$ over BINGO simulations, since the constraints from clean 21\,cm simulations rely on the same log-normal realisations. 
This second point suggests that the better constraining power of the APS for BINGO simulations could also be a consequence of a more severe impact of the contaminants over the MFs predictions compared to the APS. 
The impact of each contaminant to the summary statistics is assessed later in this section, where we get back to this discussion.

\begin{table*}
    \linespread{1.4}
    \selectfont
    \centering
    \caption{Results of the accuracy metrics evaluating the performance of the NNs trained over the MFs, the APS, and their combination, $V_k$+$C_\ell$, in predicting \{$\Omega_c, h$\} parameters. All the metrics are calculated from the test data sets; the RMSE is also estimated from the training data sets, appearing in parentheses. Given the sampling interval of the parameters (Table \ref{tab:parameters_range}), we remember that RMSE values around 0.025 and 0.030 for $\Omega_c$ and $h$, respectively, indicate inaccuracy of the predictions. The first three parts of the table show the resulting metrics from analysing the clean cosmological signal (with the beam effect; 21\,cm), these same mocks contaminated with thermal noise (21\,cm+WN), and the BINGO simulations (21\,cm+WN+FG). The last part of the table corresponds to analyses of the same clean 21\,cm realisations, but considering a larger sky coverage, $f_{\it sky} = 0.52$. See text for details.}
    \begin{tabular}{cccccccccc}
    \toprule
    \hline
    \multirow{2}{4em}{Parameter} & \multicolumn{3}{c}{$V_k$} & \multicolumn{3}{c}{$C_\ell$} & \multicolumn{3}{c}{$V_k$+$C_\ell$} \\
    \cline{2-4} \cline{5-7} \cline{8-10} 
     & $\langle \sigma \rangle$ & slope & RMSE & $\langle \sigma \rangle$ & slope & RMSE & $\langle \sigma \rangle$ & slope & RMSE \\  
    \hline
    & \multicolumn{9}{c}{21\,cm} \\
    \hline
        $\Omega_c$     & 0.007 & 0.874 & 0.014 (0.009) & 0.006 & 0.890 & 0.010 (0.009) & 0.007 & 0.910 & 0.009 (0.008) \\ 
        \hline
        $h$            & 0.006 & 0.964 & 0.010 (0.008) & 0.006 & 0.954 & 0.010 (0.008) & 0.006 & 0.943 & 0.009 (0.007)  \\
    \hline
    & \multicolumn{9}{c}{21\,cm + WN} \\
    \hline
        $\Omega_c$     & 0.010 & 0.524 & 0.020 (0.019) & 0.009 & 0.812 & 0.012 (0.012) & 0.008 & 0.808 & 0.012 (0.011) \\ 
        \hline
        $h$            & 0.007 & 0.833 & 0.014 (0.013) & 0.007 & 0.901 & 0.011 (0.010) & 0.007 & 0.922 & 0.011 (0.009)  \\
    \hline
    & \multicolumn{9}{c}{21\,cm + WN + FG (BINGO simulations)} \\
    \hline
        $\Omega_c$     & 0.010 & 0.533 & 0.020 (0.019) & 0.009 & 0.827 & 0.013 (0.011) & 0.009 & 0.784 & 0.013 (0.012) \\ 
        \hline
        $h$            & 0.008 & 0.854 & 0.015 (0.014) & 0.008 & 0.898 & 0.011 (0.010) & 0.007 & 0.890 & 0.011 (0.010)  \\
    \hline
    & \multicolumn{9}{c}{21\,cm @ $f_{\it sky} = 0.52$} \\
    \hline
        $\Omega_c$     & 0.003 & 0.967 & 0.004 (0.004) & 0.004 & 0.878 & 0.007 (0.007) & 0.004 & 0.968 & 0.004 (0.004) \\ 
        \hline
        $h$            & 0.003 & 0.968 & 0.004 (0.004) & 0.003 & 0.929 & 0.006 (0.006) & 0.003 & 0.968 & 0.004 (0.004)  \\
    \hline
    \bottomrule
    \end{tabular} \label{tab:2parameters}
\end{table*}

\begin{table}
    \linespread{1.4}
    \selectfont
    \centering
    \caption{Area of the 1$\sigma$ error ellipse for each summary statistic relative to that from the combination $V_k$+$C_\ell$, considering the different parameters planes for \{$\Omega_c, h$\} and \{$\Omega_c, h, w_0, w_a$\} constraints. The columns 2 to 4 show results from analysing the clean 21\,cm realisations, and columns 5 to 7 are obtained from BINGO simulations. }
    \begin{tabular}{ccccccc}
    \toprule
    \hline
    \multirow{2}{4em}{Parameter space} & \multicolumn{3}{c}{21\,cm} & \multicolumn{3}{c}{21\,cm+WN+FG} \\
    \cline{2-7}
    & $V_k$+$C_\ell$ & $C_\ell$ & $V_k$ & $V_k$+$C_\ell$ & $C_\ell$ & $V_k$ \\
    \hline
    \multicolumn{7}{c}{2 parameters constraint} \\
    \hline
    $\Omega_c - h$    & 1 & 1.194 & 1.188 & 1 & 1.267 & 1.686 \\ 
    \hline
    \multicolumn{7}{c}{4 parameters constraint}  \\   
    \hline
    $\Omega_c - h$    & 1. & 1.337 & 1.137 & 1. & 1.053 & 1.385 \\
    \hline
    $\Omega_c - w_0$  & 1. & 1.224 & 1.329 & 1. & 1.001 & 1.909 \\
    \hline
    $h - w_0$         & 1. & 1.016 & 1.315 & 1. & 1.017 & 1.352 \\
    \hline
    $\Omega_c - w_a$  & 1. & 1.084 & 1.138 & 1. & 1.100 & 1.301 \\
    \hline
    $h - w_a$         & 1. & 0.970 & 1.052 & 1. & 1.124 & 1.062 \\
    \hline
    $w_0 - w_a$       & 1. & 0.978 & 1.300 & 1. & 1.148 &  1.352 \\
    \hline
    \bottomrule
    \end{tabular} \label{tab:area_MFsAPS}
\end{table}

\subsection{Parameters prediction under the CPL parameterization}  \label{sec:4parameters}

\begin{figure*}
	\includegraphics[width=0.245\textwidth]{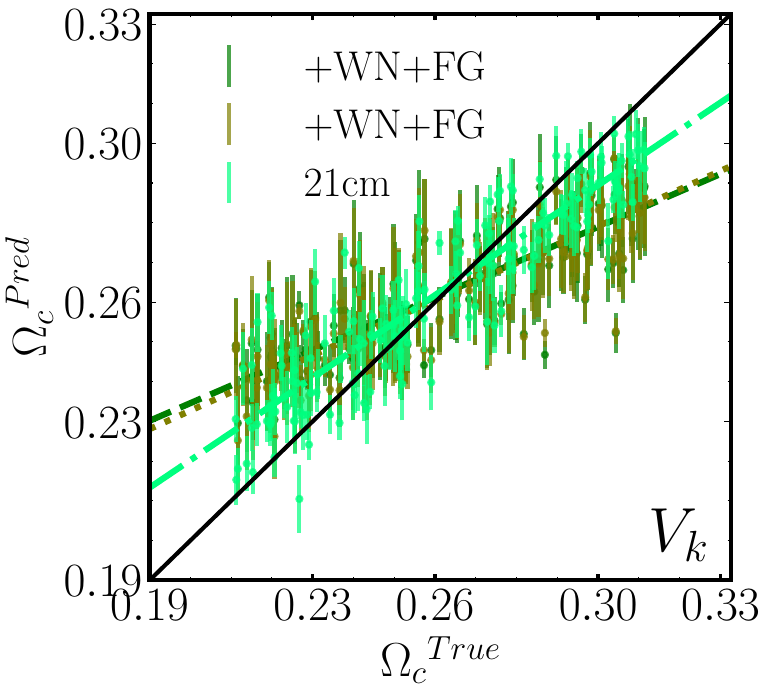}\includegraphics[width=0.245\textwidth]{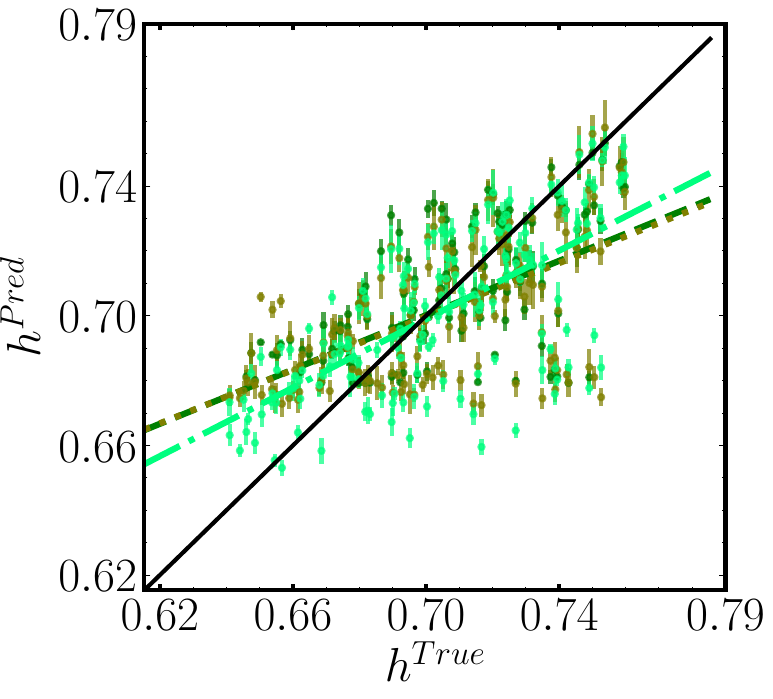}\includegraphics[width=0.26\textwidth]{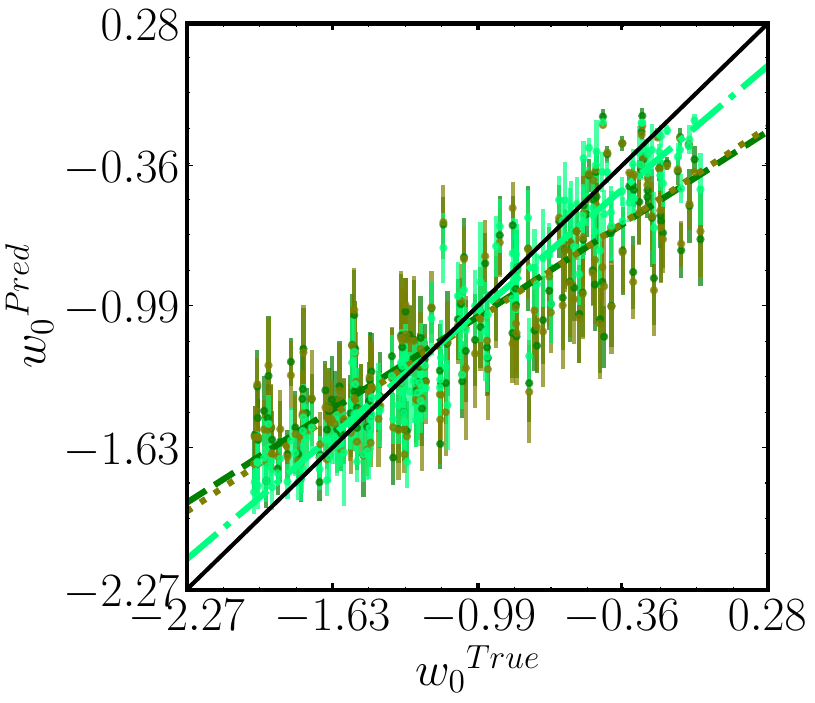}\includegraphics[width=0.26\textwidth]{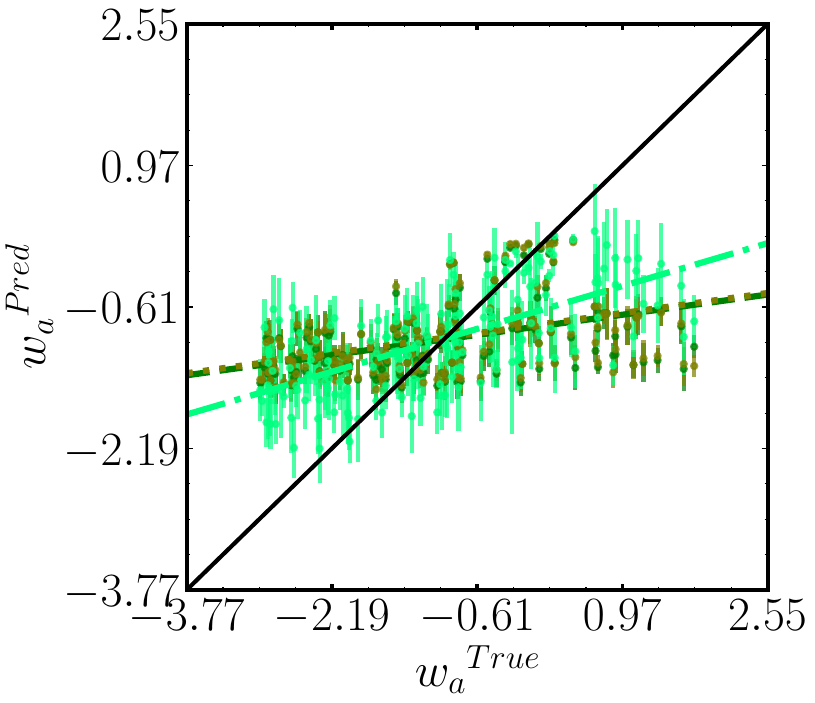}\\
	\includegraphics[width=0.245\textwidth]{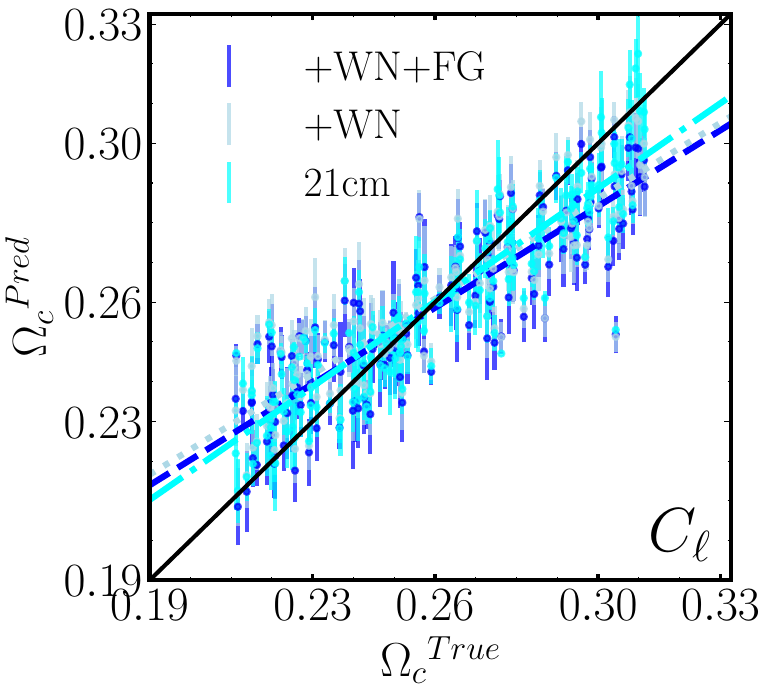}\includegraphics[width=0.245\textwidth]{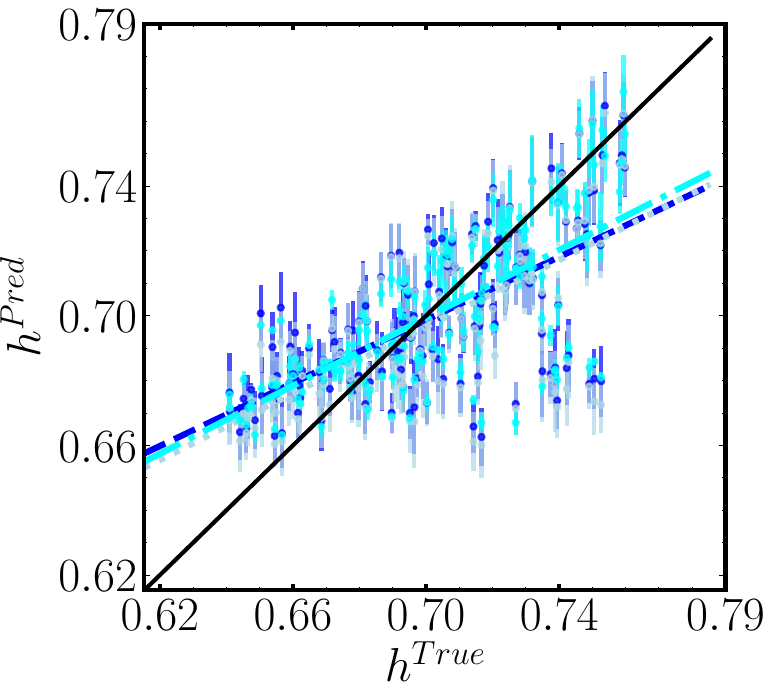}\includegraphics[width=0.26\textwidth]{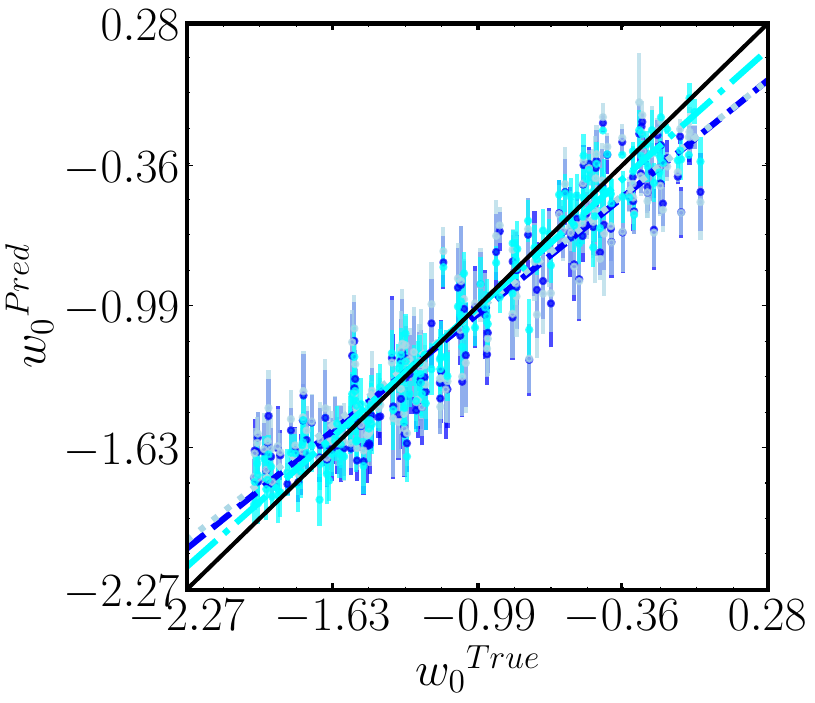}\includegraphics[width=0.26\textwidth]{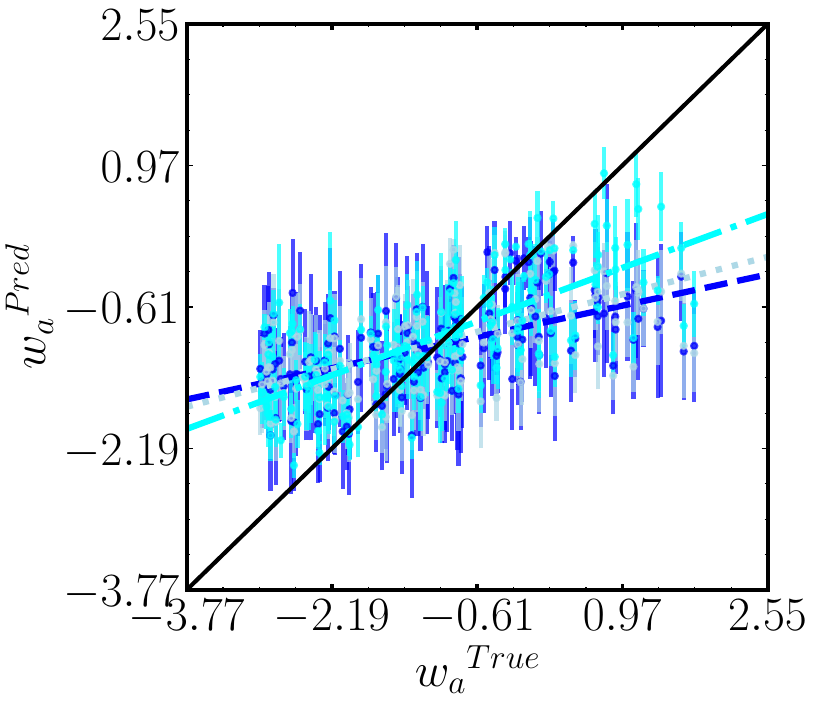}\\
	\includegraphics[width=0.245\textwidth]{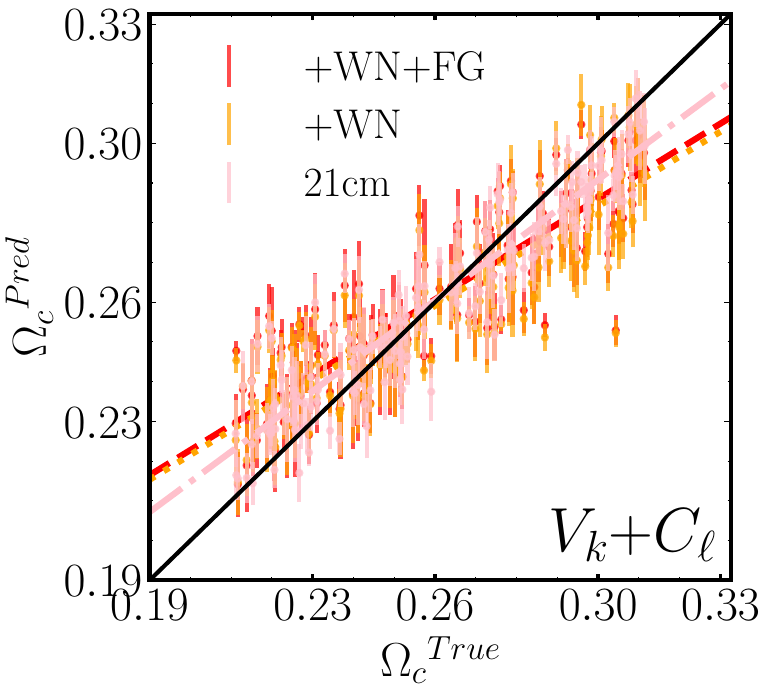}\includegraphics[width=0.245\textwidth]{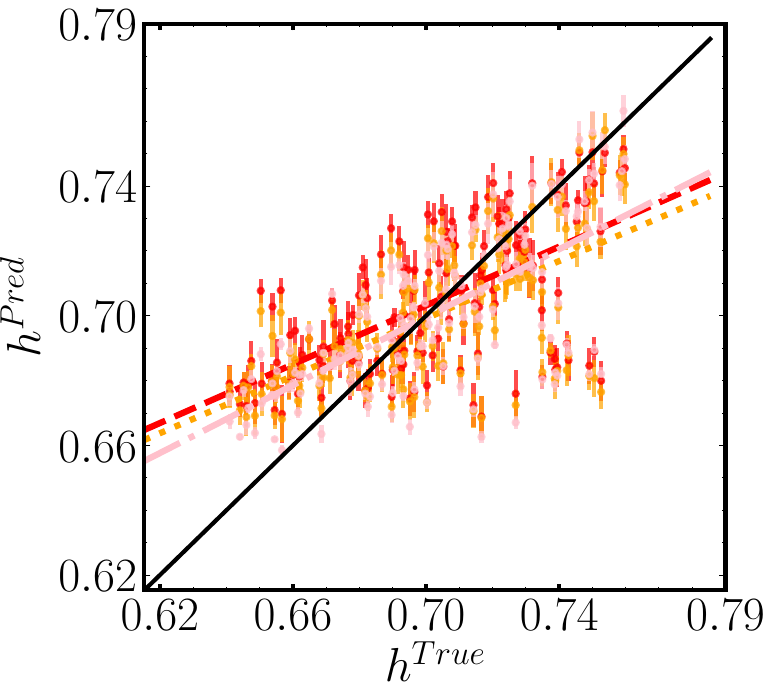}\includegraphics[width=0.26\textwidth]{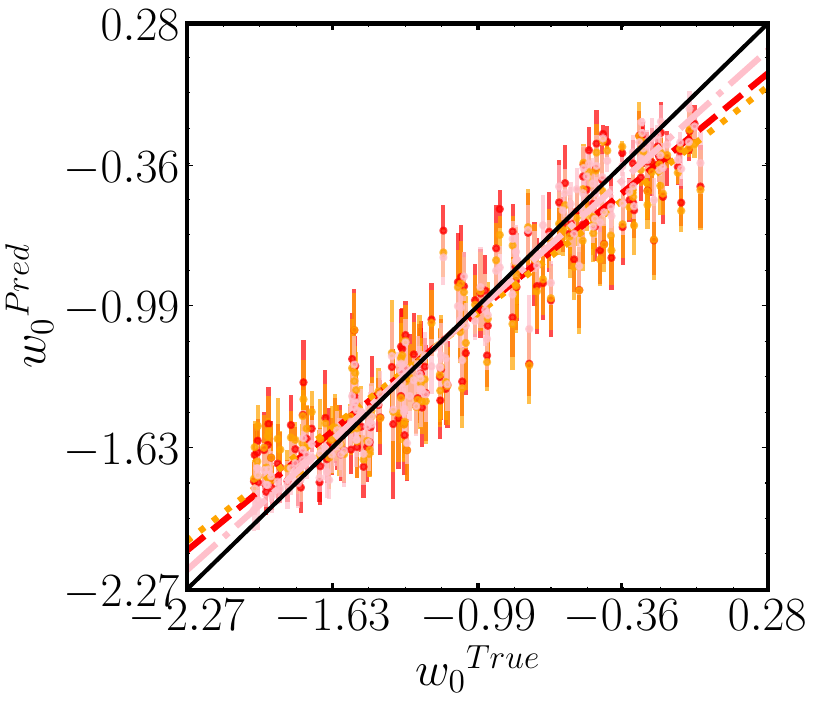}\includegraphics[width=0.26\textwidth]{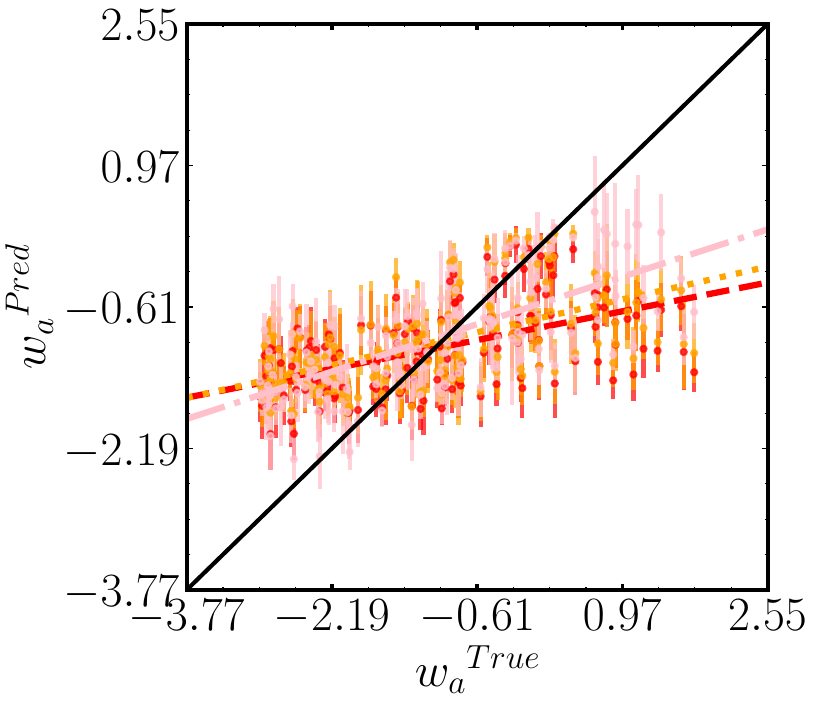}
    \caption{Same as Figure \ref{fig:2parameters}, but for case {\it (ii)}, showing predicted values for \{$\Omega_c, h, w_0, w_a$\} parameters.}
    \label{fig:4parameters}
\end{figure*}

Reproducing the same analyses as discussed in the previous subsection (same number of cosmologies and realisations), now considering the four-parameter space, case {\it (ii)} \{$\Omega_c, h, w_0, w_a$\}, we find the predicted values as shown in Figure \ref{fig:4parameters}. 
These results show a greater difficulty of the NN in mapping  the summary statistics into the cosmological parameters, which is expected given the larger parameter space. 
Also, we can see that the constraining power of the method in estimating $h$ parameter is the most affected by the inclusion of the CPL parameters, $w_0$ and $w_a$, likely due to the degeneracy among them. 
Still, from the 1$\sigma$ error bars we still see a reasonable concordance of the predictions with the true values, apart from $w_a$ parameter. 
In fact, from Figure \ref{fig:dif_MFs} in the appendix, we can see that $w_a$ is the parameter whose variation least modifies the summary statistics.  
The predictions for this parameter from both APS and MFs are clearly shifted to the mean of the respective sampling interval, artificially reducing the loss function. 
This behaviour is quantified by the slope values presented in the third part of Table \ref{tab:4parameters}. 
The same table also summarises $\langle \sigma \rangle$ and RMSE metrics resulting from the four-parameter constraints. 
Similarly to the two-parameter constraints, the metrics show that the APS outperforms the MFs in predicting all four parameters, while their combination allows only slight improvements of the metrics from using the APS alone. 
The RMSE value for $\Omega_c$, $h$, and $w_a$ obtained using $V_k$+$C_\ell$ decrease with respect to those obtained using only $C_\ell$, but slightly increase for $w_0$. 
The slope values indicate a smaller (larger) bias for $\Omega_c$ and $h$ ($w_0$ and $w_a$) parameters for $V_k$+$C_\ell$ statistics.
Again, these results can be interpreted taking into account the relationship between the two summary statistics, as we discussed in the previous section. 
Still, we should also account for the fact that the NN training and the hyperparameter optimisation are stochastic processes. 
This means that, training a NN will lead to different outputs for multiple evaluations, even fixing the set of hyperparameters, because of the random choice of some parameters of the NN algorithm and the random split of the training and validation sets. 
Also, the hyperparameter selection can lead to better optimised architectures for specific cases. 
For such reasons, a slight change in the accuracy metrics may simply indicate that there has been neither improvement nor degradation of the predictions, which would explain the few cases with slightly worse metrics from the combination $V_k$+$C_\ell$. 

The NN fed with the combination of the summary statistics provide predictions for $\Omega_c, h$, $w_0$ and $w_a$ parameters with accuracy of RMSE $\sim 0.017, 0.025$, $0.243$, and $1.096$, respectively, which gives, for the first three, 6.4\%, 3.7\%, and 24.3\% fractional error about their fiducial values (Table \ref{tab:parameters_range}). 
We note that a RMSE value close to half of the sampling interval, namely, 0.5 and 1.375, for $w_0$ and $w_a$ parameters, respectively, would indicate inaccuracy of the method in predicting these parameters; $w_a$ predictions have the higher RMSE value relative to the sampling interval.
In fact, under the CPL parameterization, the Fisher matrix analysis by \citet{2021/costa_BINGO-forecast} also finds $w_a$ parameter to be the most difficult to constrain, obtaining an error amplitude of 2.8 (1.2), for BINGO (BINGO+Planck). 
For $h$ and $w_0$ parameters\footnote{The fiducial values employed in \citet{2021/costa_BINGO-forecast} are $h = 0.6732$, $w_0 = -1$, and $w_a = 0$.} they find fractional errors of 20\% (2.9\%) and 55\% (30\%), respectively. 
Again, although a comparison between these results and ours is not straightforward, the significantly lower error estimates found here still indicates our methodology as very promising, in particular to improve Planck constraint on $w_0$. 

\begin{figure}
	\includegraphics[width=\columnwidth]{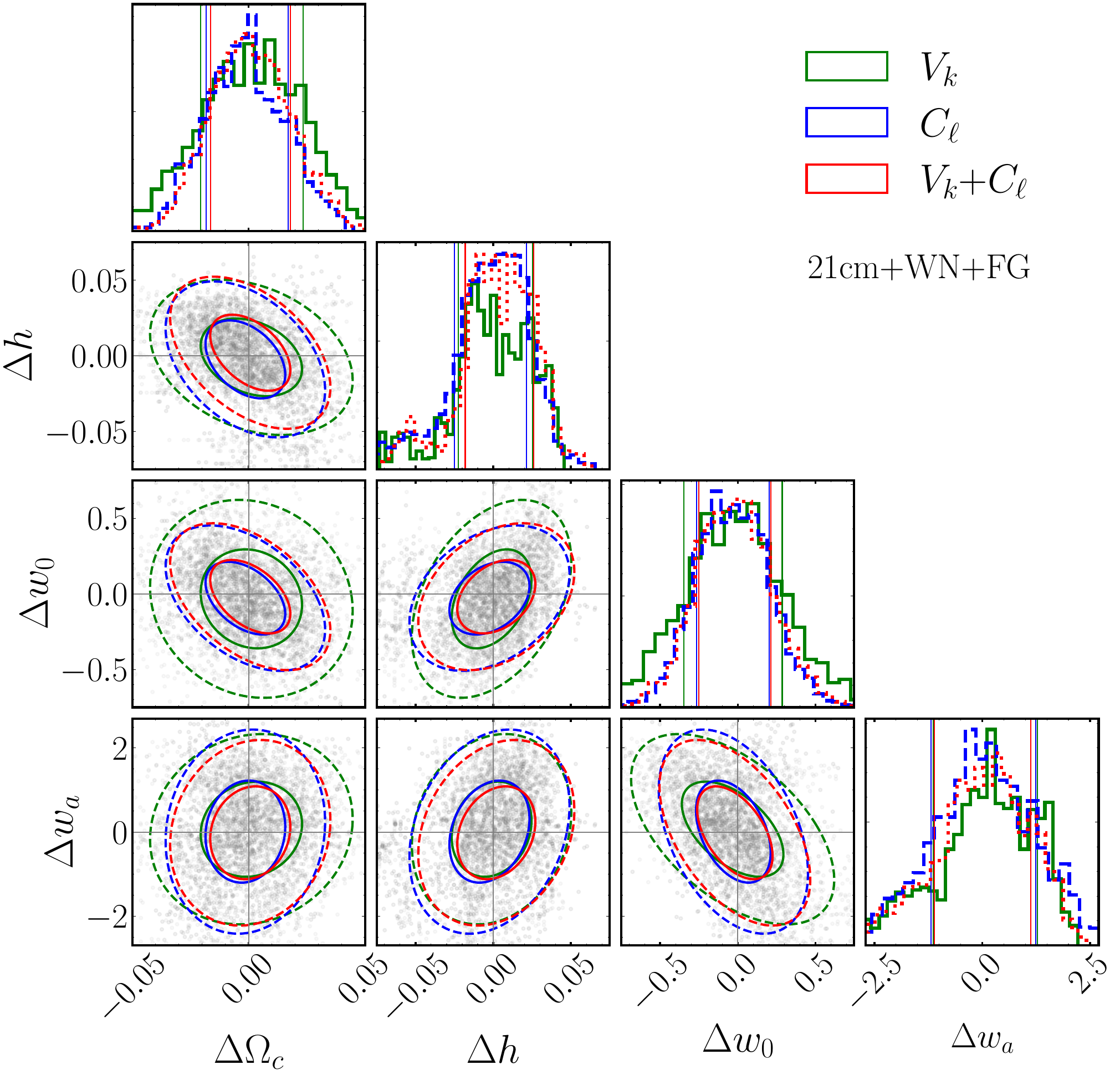}
    \caption{Same as Figure \ref{fig:triangle_2params}, but for case {\it (ii)} \{$\Omega_c, h, w_0, w_a$\} parameters.}
    \label{fig:triangle_4params}
\end{figure}

From the error ellipses shown in Figure \ref{fig:triangle_4params}, we find that, for the $\Omega_c - h$ plane, the area of the 1$\sigma$ contour from the predictions obtained using the APS is reduced by $\sim 5\%$ when the summary statistics are combined,  $V_k$+$C_\ell$, as shown in Table \ref{tab:area_MFsAPS}. 
For $\Omega_c - w_0$ and $h - w_0$ planes, the 1$\sigma$ error contours shrank by $\sim 0.1\%$ and $\sim 2\%$, respectively. 
The combination $V_k$+$C_\ell$ shrank the error contours on the $\Omega_c - w_a$, $h - w_a$, and $w_0 - w_a$ planes, obtained using the APS alone, by $\sim 10$\%, $\sim 12$\%, and $\sim 15$\%, respectively. 

Moreover, similarly to the findings from the $\Lambda$CDM scenario, when analysing the clean 21\,cm simulations the individual results from the two summary statistics (Table \ref{tab:4parameters}) are quite consistent for $h$ and $\Omega_b$ parameters, while their ellipse error show the MFs with better constraining power than the APS. 
For the other parameters and ellipse errors, the APS seems to outperform the MFs also when no contaminant signal is accounted for.

\begin{table*}
    \linespread{1.4}
    \selectfont
    \centering
    \caption{Same as Table \ref{tab:2parameters}, but for \{$\Omega_c, h, w_0, w_a$\} parameters predictions. We remind that RMSE values around 0.025, 0.030, 0.5, and 1.375 for $\Omega_c$, $h$, $w_0$, and $w_a$, respectively, indicate inaccuracy of the predictions, since the 1$\sigma$ error would amount to half of the sampling interval (Table \ref{tab:parameters_range}). }
    \begin{tabular}{cccccccccc}
    \toprule
    \hline
    \multirow{2}{4em}{Parameter} & \multicolumn{3}{c}{$V_k$} & \multicolumn{3}{c}{$C_\ell$} & \multicolumn{3}{c}{$V_k$+$C_\ell$} \\
    \cline{2-4} \cline{5-7} \cline{8-10} 
     & $\langle \sigma \rangle$ & slope & RMSE & $\langle \sigma \rangle$ & slope & RMSE & $\langle \sigma \rangle$ & slope & RMSE \\  
    \hline
    & \multicolumn{9}{c}{21\,cm} \\
    \hline
        $\Omega_c$ & 0.006 & 0.692 & 0.015 (0.016) & 0.006 & 0.713 & 0.015 (0.015) &  0.007 & 0.761 & 0.014 (0.014) \\ 
        \hline
        $h$        & 0.004 & 0.528 & 0.024 (0.024) & 0.004 & 0.524 & 0.023 (0.023) & 0.003 & 0.524 & 0.023 (0.024)  \\
        \hline
        $w_0$      & 0.125 & 0.872 & 0.214 (0.210) & 0.098 & 0.913 & 0.168 (0.159) & 0.103 & 0.919 & 0.166 (0.151)  \\
        \hline
        $w_a$      & 0.337 & 0.303 & 1.023 (1.036) & 0.313 & 0.380 & 0.967 (0.960) & 0.311 & 0.335 & 1.001 (1.005)  \\
    \hline
    & \multicolumn{9}{c}{21\,cm + WN} \\
    \hline
        $\Omega_c$ & 0.010 & 0.464 & 0.021 (0.021) & 0.007 & 0.633 & 0.017 (0.016) &  0.007 & 0.626 & 0.017 (0.017) \\ 
        \hline
        $h$        & 0.005 & 0.412 & 0.026 (0.026) & 0.009 & 0.514 & 0.026 (0.025) & 0.005 & 0.443 & 0.025 (0.024)  \\
        \hline
        $w_0$      & 0.153 & 0.679 & 0.328 (0.341) & 0.137 & 0.813 & 0.246 (0.237) & 0.133 & 0.803 & 0.240 (0.221)  \\
        \hline
        $w_a$      & 0.159 & 0.142 & 1.138 (1.168) & 0.436 & 0.265 & 1.102 (0.931) & 0.336 & 0.232 & 1.092 (1.065)  \\
    \hline
    & \multicolumn{9}{c}{21\,cm + WN + FG (BINGO simulations)} \\
    \hline
        $\Omega_c$ & 0.010 & 0.444 & 0.022 (0.021) & 0.008 & 0.638 & 0.017 (0.017) &  0.008 & 0.632 & 0.017 (0.016) \\ 
        \hline
        $h$        & 0.002 & 0.419 & 0.026 (0.027) & 0.009 & 0.486 & 0.026 (0.025) & 0.006 & 0.454 & 0.025 (0.024)  \\
        \hline
        $w_0$      & 0.155 & 0.656 & 0.329 (0.339) & 0.142 & 0.827 & 0.242 (0.230) & 0.143 & 0.843 & 0.243 (0.221)  \\
        \hline
        $w_a$      & 0.167 & 0.143 & 1.134 (1.163) & 0.609 & 0.221 & 1.214 (1.181) & 0.286 & 0.203 & 1.096 (1.096)  \\
    \hline
    & \multicolumn{9}{c}{21\,cm @ $f_{\it sky} = 0.52$} \\
    \hline
        $\Omega_c$ & 0.005 & 0.805 & 0.012 (0.013) & 0.005 & 0.797 & 0.012 (0.011) &  0.004 & 0.866 & 0.011 (0.012) \\ 
        \hline
        $h$        & 0.003 & 0.494 & 0.022 (0.023) & 0.003 & 0.587 & 0.022 (0.022) & 0.005 & 0.612 & 0.020 (0.020)  \\
        \hline
        $w_0$      & 0.087 & 0.935 & 0.145 (0.140) & 0.060 & 0.954 & 0.119 (0.115) & 0.069 & 0.963 & 0.129 (0.124)  \\
        \hline
        $w_a$      & 0.264 & 0.458 & 0.893 (0.917) & 0.285 & 0.517 & 0.840 (0.816) & 0.240 & 0.464 & 0.895 (0.903) \\
    \hline
    \bottomrule
    \end{tabular} \label{tab:4parameters}
\end{table*}

\subsection{Impact of individual systematic effects}  \label{sec:sensitivity_contaminants}

Here we evaluate how \{$\Omega_c, h$\} and \{$\Omega_c, h, w_0, w_a$\} parameters predictions are impacted by the presence of noise and foreground residual individually by adding one at a time to the cosmological signal. 
The different colours appearing in each panel of Figures \ref{fig:2parameters} and \ref{fig:4parameters} represent a type of simulation, namely, the 21\,cm only (with BINGO beam), including the thermal noise, and, along with it, adding the foreground residual. 
The metrics obtained from all these type of simulations and summary statistic are summarised in Tables \ref{tab:2parameters} and \ref{tab:4parameters}. 
These results indicate that the thermal noise causes the greatest impact on the prediction, while the presence of foreground residual, although not negligible, leads to a much less degradation of the predictions. 
For the RMSE metric, for example, considering results for $V_k$+$C_\ell$, in case {\it (i)} we find 0.009 error for both $\Omega_c$ and $h$ parameters for clean 21\,cm mocks, which increases by 30\% and 18\%, respectively, when adding noise and another 8\% and 4\% when including also the foreground residual. 
For case {\it (ii)}, the presence of thermal noise leads to RMSE values 21\%, 9\%, 44\%, and 9\% greater, for $\Omega_c, h, w_0$, and $w_a$ parameters, with respect to that obtained from predictions over clean 21\,cm mocks, while including the foreground residual has negligible impact (less than 1\%). 
As can be seen from these values, the $w_0$ parameter is the most affected by the noise contamination. 

\begin{figure}
	\includegraphics[width=\columnwidth]{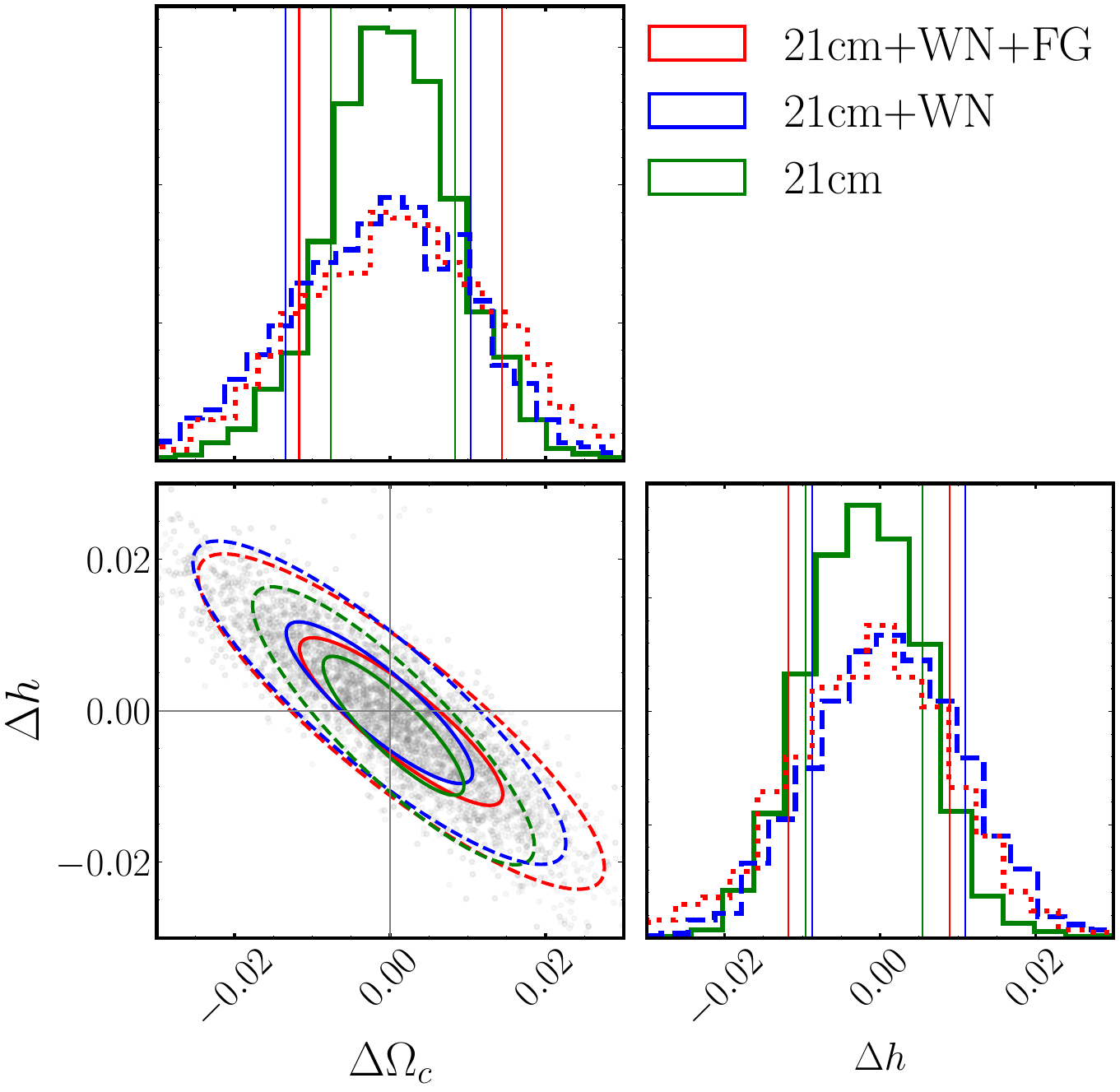}
    \caption{Same as Figure \ref{fig:triangle_2params}, but showing results from using different simulations: the clean 21\,cm simulations (21\,cm), after including thermal noise to them (21\,cm + WN), and adding foreground residual along with the noise (21\,cm + WN + FG). All predictions shown here results from the combination $V_k$+$C_\ell$.}
    \label{fig:triangle_2parameters_contaminants}
\end{figure}

\begin{figure}
	\includegraphics[width=\columnwidth]{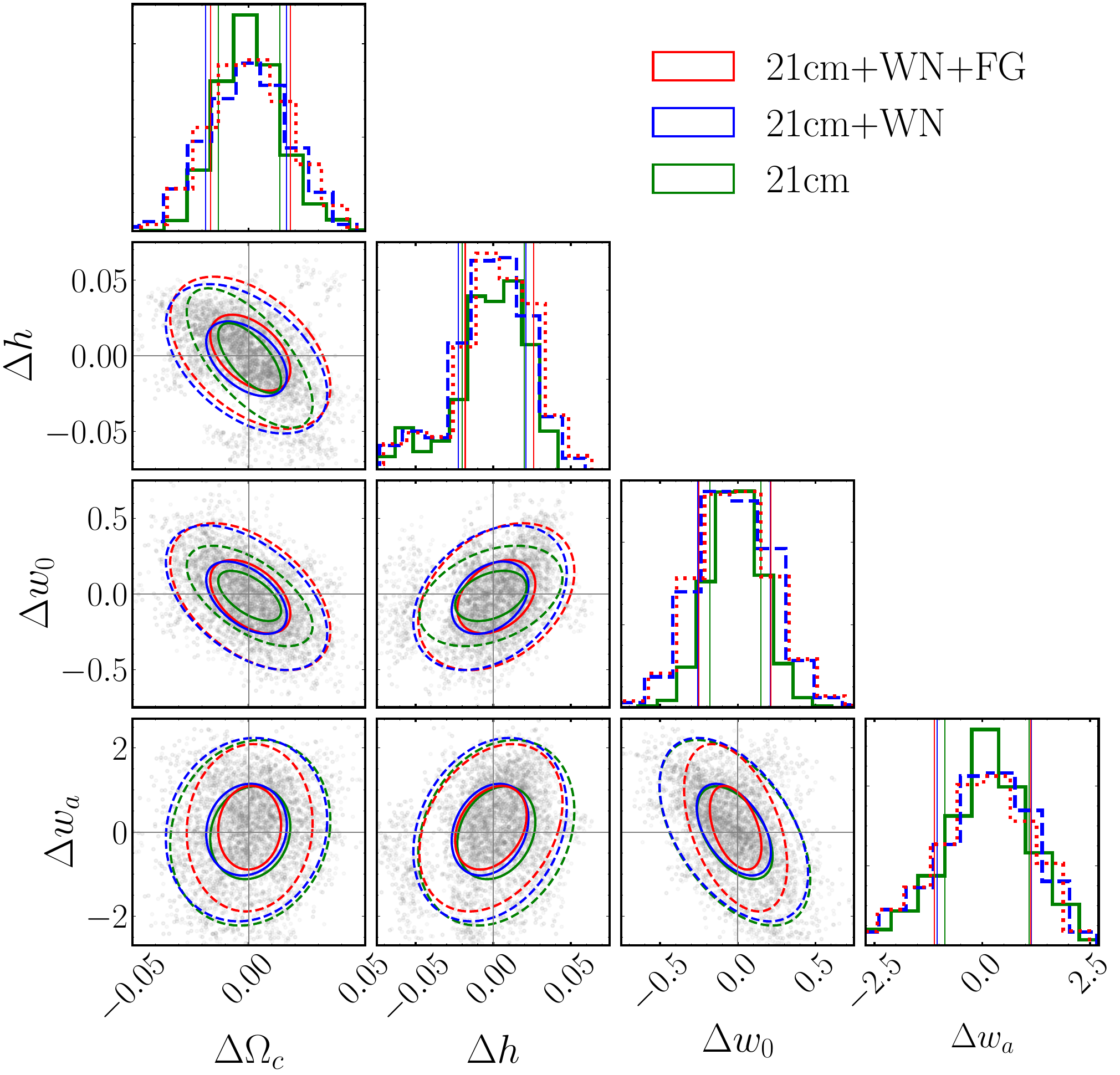}
    \caption{Same as Figure \ref{fig:triangle_2parameters_contaminants}, but for case {\it (ii)}, showing predicted values for \{$\Omega_c, h, w_0, w_a$\} parameters. }
    \label{fig:triangle_4parameters_contaminants}
\end{figure}

The comparison of the error ellipses from each type of simulation, using $V_k$+$C_\ell$ statistics, is shown in Figures \ref{fig:triangle_2parameters_contaminants}, for case {\it (i)} \{$\Omega_c, h$\}, and \ref{fig:triangle_4parameters_contaminants}, for case {\it (ii)} \{$\Omega_c, h, w_0, w_a$\}. 
Table \ref{tab:area_contaminants} shows the impact on the 1$\sigma$ area when accounting for the thermal noise and, along with it, the foreground residual (BINGO simulations). 
For all planes of parameters pairs is evident the significant impact of the thermal noise and the almost negligible effect introduced by the foreground residual. 

Evaluating the summary statistics individually, we also observe the thermal noise as the most important contaminant, while the MFs seem to be the most impacted by it. 
All metrics from predictions for case {\it (i)} show this same behaviour for both cosmological parameters. 
In this case, the RMSE calculated for the MFs increases by 42 and 40\% for $\Omega_c$ and $h$ parameters, respectively, due to the inclusion of thermal noise, while these percentages are 20 and 10\% using the $C_\ell$.
For case {\it (ii)} the increase in the RMSE due to thermal noise are 40, 8, 53, and 11\% for $\Omega_c, h, w_0$, and $w_a$, respectively, using MFs, which, using $C_\ell$, are 46\% for $w_0$ and 13\% for the other three parameters. 
Therefore, regardless of the summary statistic or their combination, $w_0$ is the parameter most affected when taking the thermal noise into account. 

As pointed out earlier in this section, the more severe impact of the contaminants over the MFs predictions compared to the APS may be one of the reasons why the constraints on $h$-$\Omega_c$ plane by the APS are more restrictive than those by the MFs when analysing the BINGO simulations and the opposite is observed for clean 21\,cm simulations. 
It is also worth reminding that both foregrounds and thermal noise have a non-Gaussian distribution over the sky, which could also contribute for the MFs to have a less constraining power than the APS for BINGO simulations. 
Also, the distinct angular scales (or multipoles) at which the APS is calculated could also could be helping the NN to discriminate between the cosmological signal, foreground residual and thermal noise, since each of them has a characteristic dependence with the angular scale.

\begin{table}
    \linespread{1.4}
    \selectfont
    \centering
    \caption{Same as Table \ref{tab:area_MFsAPS} but using only the combination $V_k$+$C_\ell$ and showing the area of the error ellipses obtained analysing noisy 21\,cm simulation (+ WN) and BINGO simulations (+ WN + FG) relative to those from clean 21\,cm simulations.}
    \begin{tabular}{cccc}
    \toprule
    \hline
    Parameter space & 21\,cm & + WN & + WN + FG (BINGO simulations) \\
    \hline
    \multicolumn{4}{c}{2 parameters constraint} \\
    \hline
    $\Omega_c - h$    & 1 & 1.522 & 1.702 \\ 
    \hline
    \multicolumn{4}{c}{4 parameters constraint}  \\   
    \hline
    $\Omega_c - h$    & 1. & 1.625 & 1.648 \\
    \hline
    $\Omega_c - w_0$  & 1. & 1.999 & 2.053  \\
    \hline
    $h - w_0$         & 1. & 1.560 & 1.610 \\
    \hline
    $\Omega_c - w_a$  & 1. & 1.400 & 1.402 \\
    \hline
    $h - w_a$         & 1. & 1.195 & 1.234  \\
    \hline
    $w_0 - w_a$       & 1. & 1.547 & 1.574 \\
    \hline
    \bottomrule
    \end{tabular} \label{tab:area_contaminants}
\end{table}

\subsection{Robustness to foregrounds} \label{sec:robustness_foregrounds}

\begin{figure*}
	\includegraphics[width=0.245\textwidth]{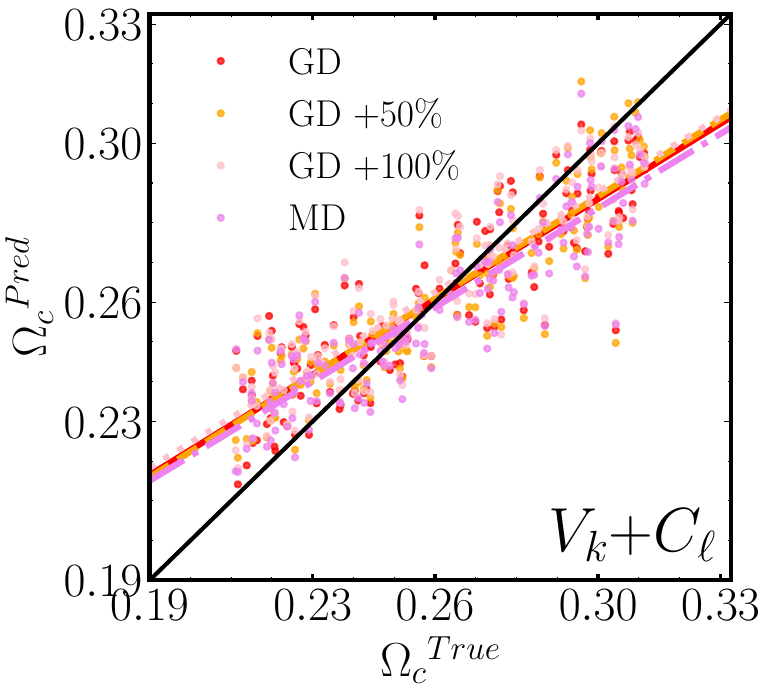}\includegraphics[width=0.245\textwidth]{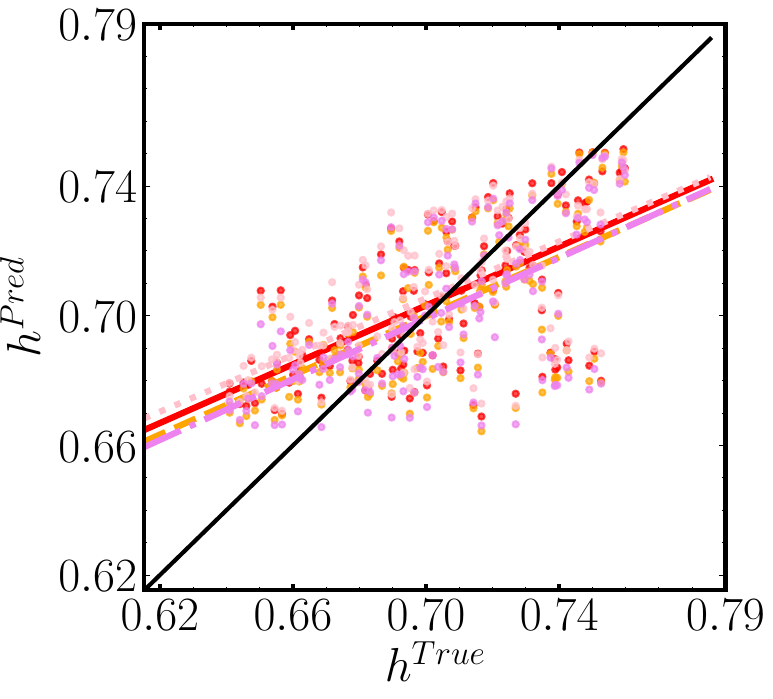}\includegraphics[width=0.26\textwidth]{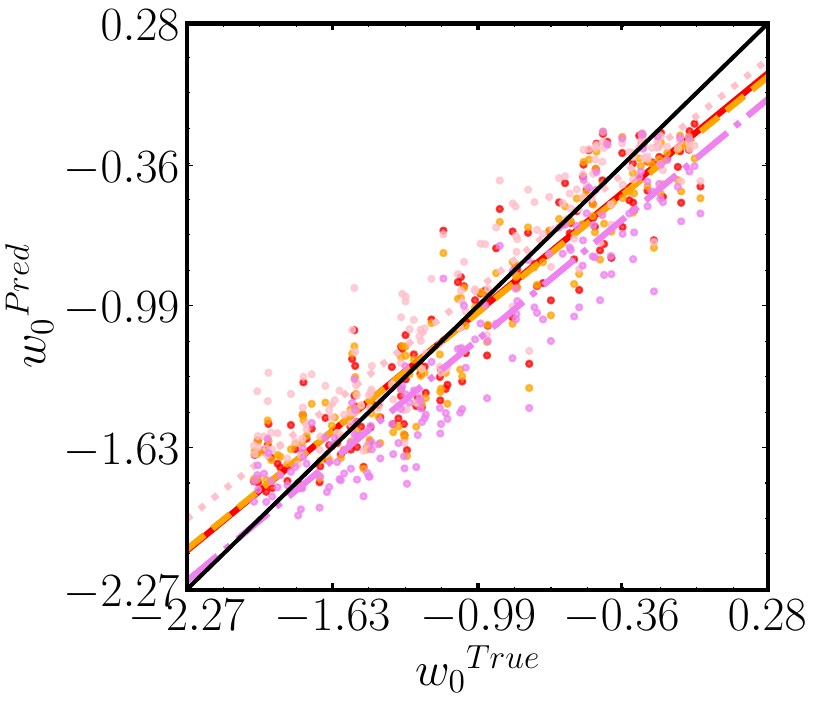}\includegraphics[width=0.26\textwidth]{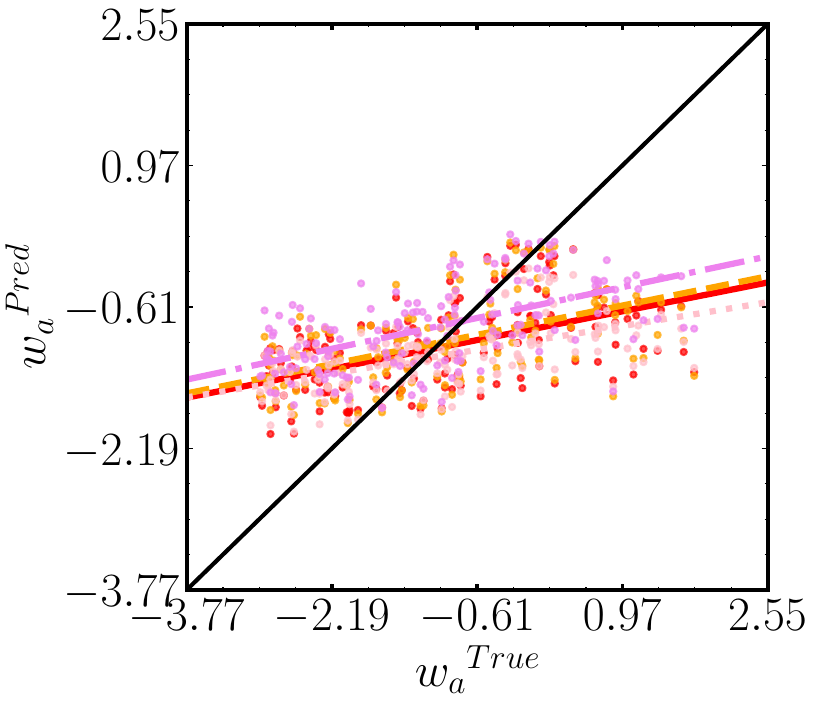}
    \caption{Same as last row of Figure \ref{fig:4parameters}, but comparing results obtained using different training data sets. 
    The red line and dots repeat +WN+FG results from Figure  \ref{fig:4parameters}, indicated here by GD \citep[synchrotron model from][]{2002/giardino}. 
    The lines and dots in orange and pink correspond to the cases using 50\% and 100\% higher residual foreground, respectively. 
    The violet line and dots correspond to the results using a different synchrotron model \citep[][MD model]{2008/miville}. 
    The error bars are not shown to help visualize the results. 
    See text for details.}
    \label{fig:4parameters_fgTest}
\end{figure*}

We also evaluate the robustness of our results from case {\it (ii)} \{$\Omega_c, h, w_0, w_a$\} to the foreground contamination. 
For this we test the previously trained NNs over three different test data sets, namely, the summary statistics obtained from simulations accounting for: (1) 50\% and (2) 100\% higher amplitude foreground contamination, by multiplying the same foreground residual maps by the factors of 1.5 and 2.0  before adding them to the cosmological signal (mimicking a non-optimal usage of {\tt GNILC}), and (3) using the foreground residual obtained as explained in section \ref{sec:foreg_clean_proc}, but using a different synchrotron emission model. 
In this last case, the synchrotron component is produced considering a power law with a non-uniform spectral index over the sky as given by the \cite{2008/miville} model (indicated by MD in Figure \ref{fig:4parameters_fgTest}), replacing the \cite{2002/giardino} model (GD in Figure \ref{fig:4parameters_fgTest}) considered for the training data set \cite[see][for details]{2022/mericia_BINGO-component_separation_II}. 
We emphasize that, to obtain the last test data set, we again apply GNILC over ten complete simulations, now accounting for the synchrotron MD model, so that the foreground residual maps are different from those used to produce the training data set. 

The results for parameters constraints using each of these three test data sets are shown in Figure \ref{fig:4parameters_fgTest}, along with the previous results (indicated by +WN+FG in the last row of Figure \ref{fig:4parameters}). 
Comparing them, we find no significant degradation of the results. 
From all three test data sets, the $\langle\sigma\rangle$ and slope metrics are highly consistent with previous predictions, presented in Table \ref{tab:4parameters} (21\,cm + WN + FG), regardless of the cosmological parameter. 
Similarly, the RMSE increases only up to 2\% for $\Omega_c$, $h$ and $w_a$. 
The larger impact is observed for $w_0$, with RMSE increasing by 11\% and 19\% for test data sets (2) and (3), respectively, which corresponds to a fractional error increasing from 24.3\% to 27.1\% and 29.1\%. 
For test data set (1), no degradation is observed from any metric or cosmological parameter. 
Therefore, even with a larger impact on $w_0$ when using a different synchrotron model, our results still corroborate the applicability of the NNs outside the training data set, showing no significant additional bias on the cosmological parameters constraints. 
This suggests our predictions to be robust against foreground contamination and our method promising to be applied to future data. 

\subsection{Sensitivity to survey area} \label{sec:sensitivity_area}

In order to evaluate how the sky coverage influences our results, we tested our methodology over a larger sky fraction, $f_{\it sky} = 0.52$. 
We chose to assume approximately the same region that SKA is intended to observe. 
We emphasise that we are not reproducing the SKA observational specifications, but only increasing the sky fraction to that expected to be covered by it, i.e., still using the same 21\,cm mocks generated in BINGO frequency range. 
The cut sky mask in this case is constructed assuming the observed region in the declination range $(-75^\circ, 28^\circ)$. 
In addition to this cut, we remove the region where Galactic foreground emission would contribute and apodize the mask, following the same procedure described in Subsection \ref{sec:observations}. 
Although we apply a Galactic cut, this test is performed only over the clean 21\,cm simulations, including the 40 arcmin beam effect. 

The last part of Tables \ref{tab:2parameters} and \ref{tab:4parameters} show the accuracy metrics, for cases {\it (i)} \{$\Omega_c, h$\} and {\it (ii)} \{$\Omega_c, h, w_0, w_a$\} parameters predictions, respectively, evaluating the performance of the method over a larger sky area. 
Compared to the results shown in the first part of these tables, one can clearly see the improvement in all the metrics when using a larger area, as expected. 
Increasing the $f_{\it sky}$ from 0.09 to 0.52, the RMSE from case {\it (i)} can decrease by a factor 2.25, for both parameters, when using $V_k$+$C_\ell$, with an even more significant improvement when using only the MFs. 
When constraining two cosmological parameters, the improvement on predictions from MFs are high enough to make them outperform the $C_\ell$. 
In fact, this is an advantageous aspect, since the MFs are the summary statistic most affected by thermal noise, and indicates that using a larger fraction of the sky may allow the combination $V_k$+$C_\ell$ to provide better results compared to individual statistics when accounting for the contaminant signals. 
Similarly, although the improvements are not so expressive for case {\it (ii)}, all the metrics improve when using a larger fraction of the sky (for both summary statistics and their combination), with smaller RMSE values by a fact at most 1.3 for $\Omega_c$ and $w_0$ using $V_k$+$C_\ell$. 

\begin{figure}
	\includegraphics[width=\columnwidth]{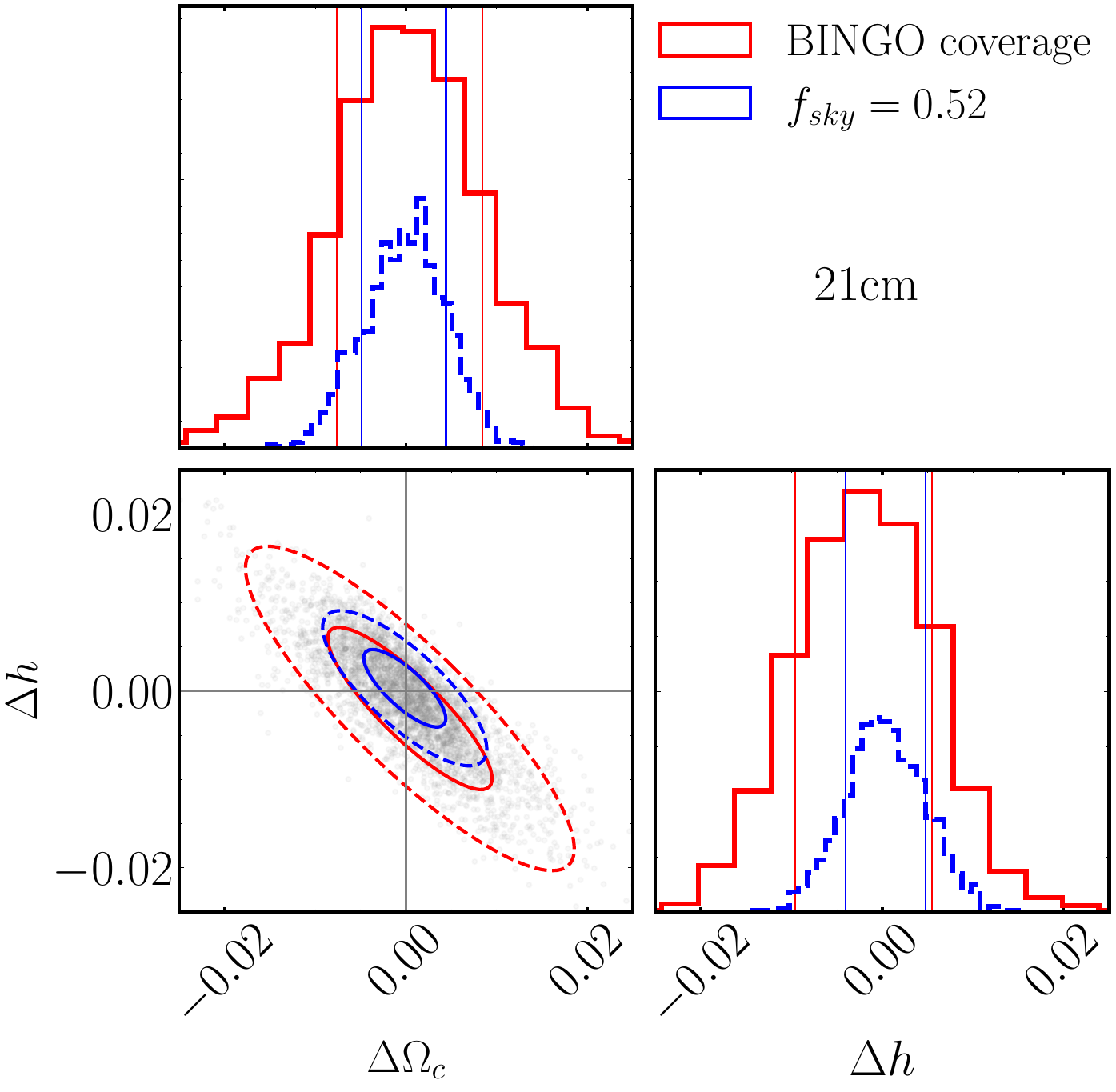}
    \caption{Error ellipses for the \{$\Omega_c, h$\} parameter space, showing the distribution of the difference of predicted and true values, as in Figure \ref{fig:triangle_2params}, but comparing two different sky coverages, $f_{\it sky} = 0.09$ (BINGO coverage) and $0.52$. Both analyses use clean 21\,cm simulations and the combination of the summary statistics, $V_k$+$C_\ell$. }
    \label{fig:triangle_fsky_2parameters}
\end{figure}

\begin{figure}
	\includegraphics[width=\columnwidth]{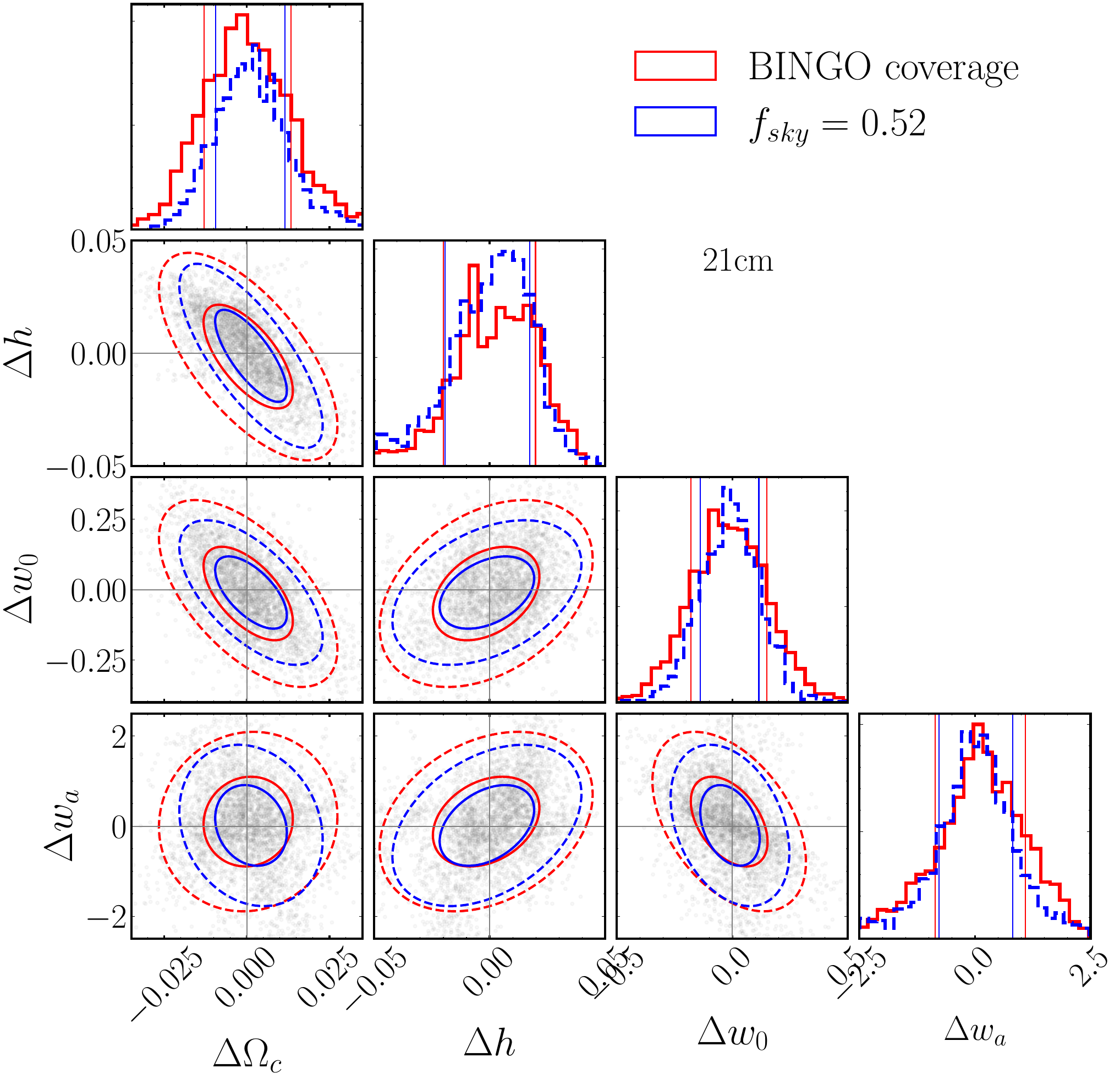}
    \caption{Same as Figure \ref{fig:triangle_fsky_2parameters}, but for case {\it (ii)}, showing predicted values for \{$\Omega_c, h, w_0, w_a$\} parameters.}
    \label{fig:triangle_fsky_4parameters}
\end{figure}

The error ellipses for $f_{\it sky}$ = 0.09 (BINGO coverage) and 0.52 are compared in Figures \ref{fig:triangle_fsky_2parameters} and \ref{fig:triangle_fsky_4parameters} for cases {\it (i)} and {\it (ii)}, respectively. 
The effect of the sky fraction over the area of the 1$\sigma$ regions are summarised in Table \ref{tab:area_fsky}, showing a significant reduction of this area from all planes when using the combination $V_k$+$C_\ell$ over a larger $f_{\it sky}$, mainly favouring $\Omega_c-h$ plane from case  {\it (i)} and $\Omega_c-w_0$ from case  {\it (ii)}. 
Besides, the results confirm the MFs outperforming the $C_\ell$ for $\Omega_c-h$ plane in clean 21\,cm maps, for both sky fractions (see columns 2 to 4 in Table \ref{tab:area_MFsAPS}) and cases {\it (i)} and {\it (ii)}. 
For case {\it (ii)}, the larger sky fraction allows the 1$\sigma$ area for $\Omega_c-h$ plane obtained using the MFs to be reduced in $\sim$22\% when combing the summary statistics, $V_k$+$C_\ell$. 
A similar reduction is also found in $\Omega_c-w_0$ plane. 
Also, using $f_{\it sky}$ = 0.52 leads to almost 5\% reduction of the 1$\sigma$ area from both $\Omega_c - w_a$ and $h - w_a$ planes using $V_k$+$C_\ell$ with respect to their individual usage, while $w_0 - w_a$ plane is slightly better constrained by the $C_\ell$ alone. 
For $f_{\it sky}$ = 0.09, as shown in Table \ref{tab:area_MFsAPS}, only $\Omega_c - w_a$ has the ellipse area reduced by the combination $V_k$+$C_\ell$. 
Such results show that increasing the sky coverage improves the predictions of all cosmological parameters, as expected, as well as allows the combination of summary statistics to provide more expressive amelioration of the predictions compared to their individual usage.

\begin{table}
    \linespread{1.4}
    \selectfont
    \centering
    \caption{Area of the 1$\sigma$ error ellipse for individual summary statistics relative to that from the combination $V_k$+$C_\ell$, but analysing a larger fraction of the sky, $f_{\it sky} = 0.52$ (columns 2 and 3). The last column shows the relative area of the ellipse when analysing BINGO coverage ($f_{\it sky} = 0.09$) using $V_k$+$C_\ell$. These results correspond to clean 21\,cm simulations.}
    \begin{tabular}{ccccc}
    \toprule
    \hline
    \multirow{2}{4em}{Parameter space} & \multicolumn{3}{c}{$f_{\it sky} = 0.52$} & \multicolumn{1}{c}{BINGO coverage} \\
    \cline{2-5}
    & $V_k$+$C_\ell$ & $C_\ell$ & $V_k$ & $V_k$+$C_\ell$ \\
    \hline
    \multicolumn{5}{c}{2 parameters constraint} \\
    \hline
    $\Omega_c - h$    & 1 & 2.902 & 1.010 & 3.383 \\ 
    \hline
    \multicolumn{5}{c}{4 parameters constraint}  \\   
    \hline
    $\Omega_c - h$    & 1. & 1.605 & 1.223 & 1.615 \\
    \hline
    $\Omega_c - w_0$  & 1. & 1.218 & 1.254 & 1.659 \\
    \hline
    $h - w_0$         & 1. & 1.012 & 1.233 & 1.479 \\
    \hline
    $\Omega_c - w_a$  & 1. & 1.048 & 1.101 & 1.404 \\
    \hline
    $h - w_a$         & 1. & 1.046 & 1.084 & 1.296 \\
    \hline
    $w_0 - w_a$       & 1. & 0.831 & 1.132 & 1.362 \\
    \hline
    \bottomrule
    \end{tabular} \label{tab:area_fsky}
\end{table}

\subsection{Sensitivity to redshift ranges} \label{sec:sensitivity_zbins}

\begin{figure}
	\includegraphics[width=0.9\columnwidth]{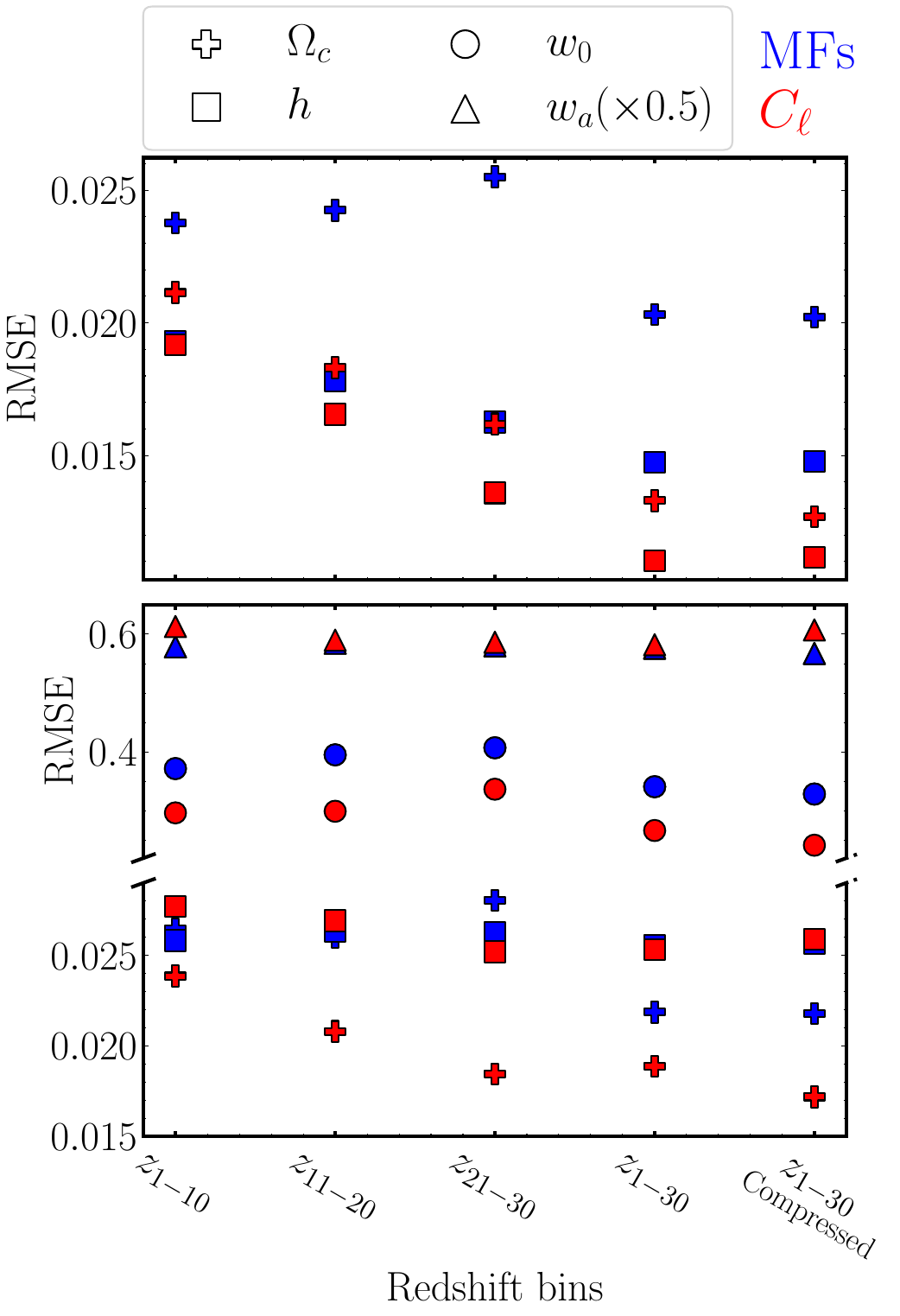}
    \caption{Sensitivity of the parameters predictions with the redshift. The upper and lower panels show the RMSE metric obtained from cases {\it (i)} and {\it (ii)}, respectively, as a function of the range of redshift bins from which the summary statistics are calculated and used to feed the NNs. The RMSE values for $w_a$ parameter are re-scaled by a factor of 0.5. In both panels, the right most data points show the RMSE resulting from predictions performed using the summary statistics from all the 30 redshift bins averaged into 5. These results are obtained using BINGO simulations. See text for details.}
    \label{fig:rmse_zbins}
\end{figure}

We also investigate the sensitivity of each cosmological parameter to the redshift. 
For this, we perform the same training procedure described before, using as features the summary statistics from a given range of redshift. 
In addition to the whole set of 30 redshift bins, we also test using three subsets, the lowest, intermediate and highest ten redshift bins, which correspond to $z_{\small 1-10} \in [0.127, 0.234]$, $z_{\small 11-20} \in [0.234, 0.342]$, and $z_{\small 21-30} \in [0.342, 0.449]$, respectively. 
The RMSE metric for each of these tests, for cases {\it (i)} and {\it (ii)}, are presented in Figure \ref{fig:rmse_zbins} showing results for each cosmological parameter and summary statistic. 
Since using two different summary statistics (or 5, given that the MFs include the Area, Perimeter, and Genus) from 30 redshift bins constitute a very large set of features, we also evaluate the effect of compressing the summary statistics. 
This compression corresponds to a simple average of the statistics from each 6 consecutive bins, so that we have, effectively, 5 redshift bins instead of 30\footnote{We also evaluate the Principal Components Analysis (PCA) technique for such dimensionality reduction, i.e., to select only the relevant features for the training process. The results do not show improvements with respect to using the 5 compressed redshift bins.}. 
The RMSE values resulting from employing this compression is also included in Figure \ref{fig:rmse_zbins}.

Comparing the results, we can see that, for both cases {\it (i)} and {\it (ii)}, the $\Omega_c$ parameter is better predicted at lower (higher) redshifts when using the $V_k$ ($C_\ell$), with the RMSE increasing (decreasing) with redshift. 
For the $h$ parameter we find that, for both $V_k$ and $C_\ell$ in case {\it (i)}, the higher the redshift the better the predictions. 
In case {\it (ii)}, however, the predictions do not have a clear dependence with redshift, regardless of the summary statistic. 
Also, in this particular case, the combination of all redshift bins or their compression into 5 bins does not allow a very expressive improvement of the predictions. 
This behaviour confirms that when sampling also the CPL parameters, although still able to reasonably predict the $h$, as we show before, the mapping between the summary statistics and cosmological parameters is not able to keep the same efficiency as in case {\it (i)}. 

Finally, for $w_0$ parameter we find a similar behaviour for both summary statistics, with increasing RMSE metric with redshift. 
For $w_a$, in contrast, we see no clear dependence with the redshift range.
In fact, apart from $h$ and $w_a$ parameters in case {\it (ii)}, all other results shown in Figure \ref{fig:rmse_zbins} show that the better predictions are obtained using all the 30 redshift bins, or their compression. 
This motivated us to employ the compressed summary statistics in the analyses presented in previous sections. 

In addition, we can again use Figure \ref{fig:dif_MFs} to interpret these results. 
This figure shows that, although the differences observed in the MFs due to the 1\% variation of the cosmological parameter have large amplitudes at smaller redshifts, the 1$\sigma$ regions significantly decrease at higher redshifts, no longer overlapping, for the first two MFs, Area and Perimeter, and the APS. 
This means that, at higher redshifts, the cosmic variance would have a smaller impact on the predictions. 
This can help explain the improvements of $h$ parameter predictions with redshift, regardless of the summary statistic, and of $\Omega_c$ for the $C_\ell$. 
The opposite is observed from $\Omega_c$ predictions using the MFs, as well as for $w_0$ with both statistics, whose RMSE metrics increase with redshift, which indicate that the prediction of these two parameters may rely more significantly on other aspect of the summary statistics.

\section{Conclusions} \label{sec:conclusions}

We evaluate the performance of NNs in learning the relationship among summary statistics (APS and MFs) from the redshifted {\hi} emission line and cosmological parameters, preventing the use of analytical likelihoods and assumptions for their constructions. 
For this we employ a large set of 21\,cm signal simulations, spanning a grid of 800 cosmologies, generated using the {\tt FLASK} code, largely sampling the parameter space, using as a case study the BINGO telescope; our BINGO simulations account for the sky coverage and instrumental beam, besides the contamination by thermal noise and foreground residual. 
Such data set is split so that 64\%, 16\%, and 20\% of the cosmologies are used for training, validation and test, respectively.
We evaluate the constraining power of our method under two scenarios, the $\Lambda$CDM model, predicting the dark matter and Hubble parameters \{$\Omega_c, h$\} (case {\it (i)}), and the CPL parameterization, accounting also for the DE EoS parameters, i.e., predicting \{$\Omega_c, h, w_0, w_a$\} (case {\it (ii)}). 
Although we are interested in predicting, in each scenario, the aforementioned parameters, our grid of cosmologies is defined varying also $\Omega_b$, $n_s$, and $A_s$ parameters in a more restrictive range of values, given by the $1\sigma$ confidence region from Planck 2018. 
As far as we know, no other work from the literature accounts for the uncertainty of parameters other than those of interest. 
In fact, since the exact values of the cosmological parameters are not known, but only their most likely values and respective error bars, sampling other parameters of the cosmological model seems more appropriate, avoiding underestimating errors. 
The relatively fast production of our lognormal simulations allows us to better explore the parameter space.

We find that, under the $\Lambda$CDM scenario, both $\Omega_c$ and $h$ parameters can be quite well predicted by the trained NN, obtaining 4.9\% and 1.6\% fractional errors, respectively, using the combination of APS and MFs statistics. 
Such constrains, in particular for $h$, are significantly impacted by the inclusion of the CPL parameters; we find predictions for $\Omega_c$, $h$, and $w_0$ with 6.4\%, 3.7\%, and 24.3\% errors. 
The Hubble parameter is still reasonably well constrained by our method. 
Predictions of $w_a$, in contrast, have a large bias, being poorly constrained. 

Investigating the constraining power of each summary statistic individually, we find the APS outperforming the MFs in most of our tests over BINGO simulations, which can be partially explained by a possible imprecision of the three-point information of our lognormal simulations. 
Comparing such results to those obtained analysing clean 21\,cm simulation, we find that the MFs are the most impacted by the presence of these contaminants, which may help explain the less effective constraining power of the MFs.
We note that the greater impact of the presence of contaminants over the MFs predictions could possibly be associated to their non-Gaussian characteristic or even because the cosmological signal, foregrounds and thermal noise are more easily distinguishable in terms of angular scales, as given by the APS, than $\nu$ thresholds.

We assess the contribution of each contaminant signal individually to the results and find that the accuracy of the predictions is mainly affected by thermal noise, while foreground residual has a minor impact.
The robustness of our methodology and results to the foreground contamination is also evaluated by applying the trained NNs to test data sets accounting for different contributions of foreground residual, confirming that our NNs can be efficiently applied outside the training data. 
Furthermore, we find that the accuracy of the cosmological parameters predictions is sensitive to the redshift range, which is also determined by the summary statistic feeding the NN. 
The usage of the full range of redshift is confirmed to provide the better predictions. 

Using clean 21\,cm simulations, we also investigate how a four times large sky coverage than BINGO's, e.g., as the SKA footprint, can improve cosmological constrains. 
We find that, under the $\Lambda$CDM scenario, the $\Omega_c - h$ plane has its 1$\sigma$ error ellipse reduced by a factor of 3. 
Also, increasing the sky fraction seems to allow the combination of the summary statistics to provide a tighter 1$\sigma$ contour compared to their individual usage, as observed for most of the parameters planes evaluated here, in particular for $\Omega_c - h$ predictions.

Moreover, although the suboptimal cosmological simulations employed here may also have contributed to the less effective constraining power of the MFs compared to the APS, or their very similar performance in some cases, the {\tt FLASK} mocks have proved to be enough for our purpose. 
They allowed us to assess the cosmological constraining power of our method, showing the usage of NN fed by summary statistics from 21\,cm simulations to be a very promising  alternative to likelihood based analysis. 
In addition, they have also allowed to demonstrate how simple can be the combination of summary statistics to feed the NN and predict cosmological parameters, preventing technical problems commonly faced by standard analysis, as, e.g., when calculating covariance matrices. 
In fact, such simulations, although enough to evaluate the role and efficiency of the APS as features, are suboptimal and may be preventing us of exploring the full potential of the MFs, expected to carry a larger amount of information with respect to the APS due to their sensitivity to higher order information.  
Fore this reason, we believe to be able to improve our results by using, for example, N-body or hydrodynamic simulations, which we leave for future work. 
More suitable simulations are expected to allow for better predictions using the MFs and more expressive improvements from the combination with the APS and other summary statistics.

It is worth emphasising that, in addition to the 21\,cm simulations, some aspects of our analyses also deserve improvements, left for a future extension of this work. 
In particular, the improvement of BINGO simulations, making them more realistic by accounting for possible signal remaining after cleaning the $1/f$ noise \citep{2015/bigot-sazy} and the contamination by polarisation leakage \citep{2021/cunnington}. 

Even using suboptimal simulations, our results report the success of the NNs in mapping the summary statistics into the cosmological parameters. 
In particular, they indicate that future low redshift 21\,cm data should be able to provide important contributions, in special to help investigate the Hubble tension and to study the dark sector. 
Finally, we emphasise that the methodology described here is not restricted to the BINGO case, but can be explored and optimised to use different data sets, such as from other 21\,cm experiments, galaxy surveys, CMB, or even a given combination of them.

\section*{Acknowledgements}

C.P.N. thanks Serrapilheira and São Paulo Research Foundation (FAPESP; grant 2019/06040-0) for financial support. M.R. acknowledges support from the CSIC programme 'Ayuda a la Incorporaci\'on de Cient\'ificos Titulares' provided under the project 202250I159. This research made use of {\tt astropy} \citep{2018/astropy}, {\tt healpy} \citep{2019/healpy}, {\tt numpy} \citep{2011/numpy}, {\tt scipy} \citep{2020/scipy} and {\tt matplotlib} \citep{2007/matplotlib}.

\section*{Data Availability}

The data underlying this article, as well as NN models and algorithm, are available upon reasonable request to the corresponding author.
 



\bibliographystyle{mnras}
\bibliography{ClsMFsNN_forParamsConstraints.bib} 




\appendix

\section{MFs dependence with cosmological parameters} \label{appendixA}

\begin{figure*}
	\includegraphics[width=0.972\textwidth]{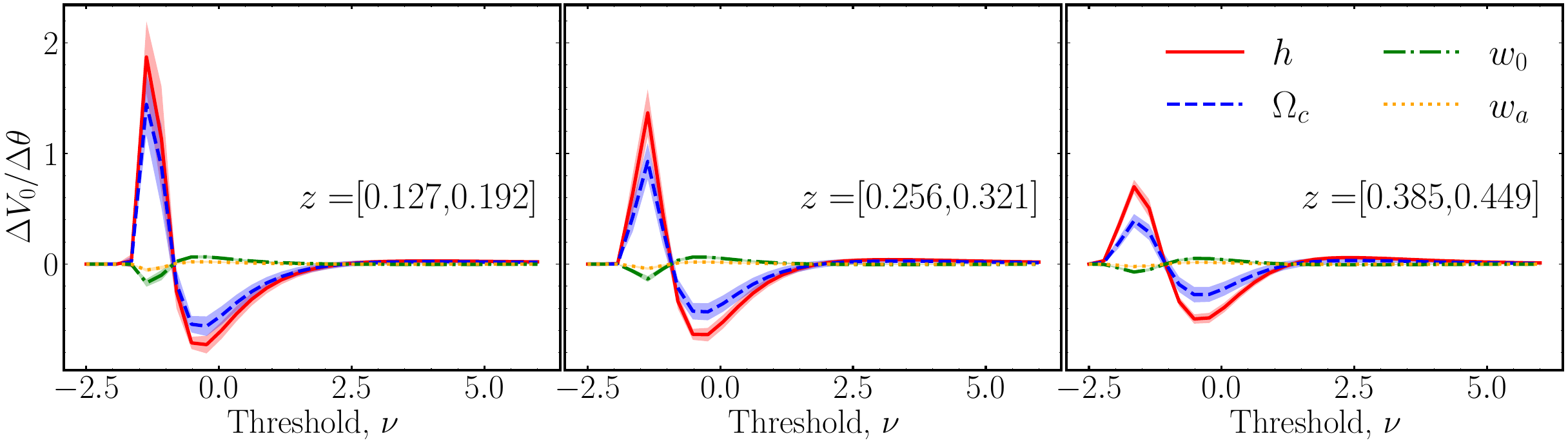}\\
    \vspace{-0.85cm}
 	\hspace{-0.6cm}\includegraphics[width=1.003\textwidth]{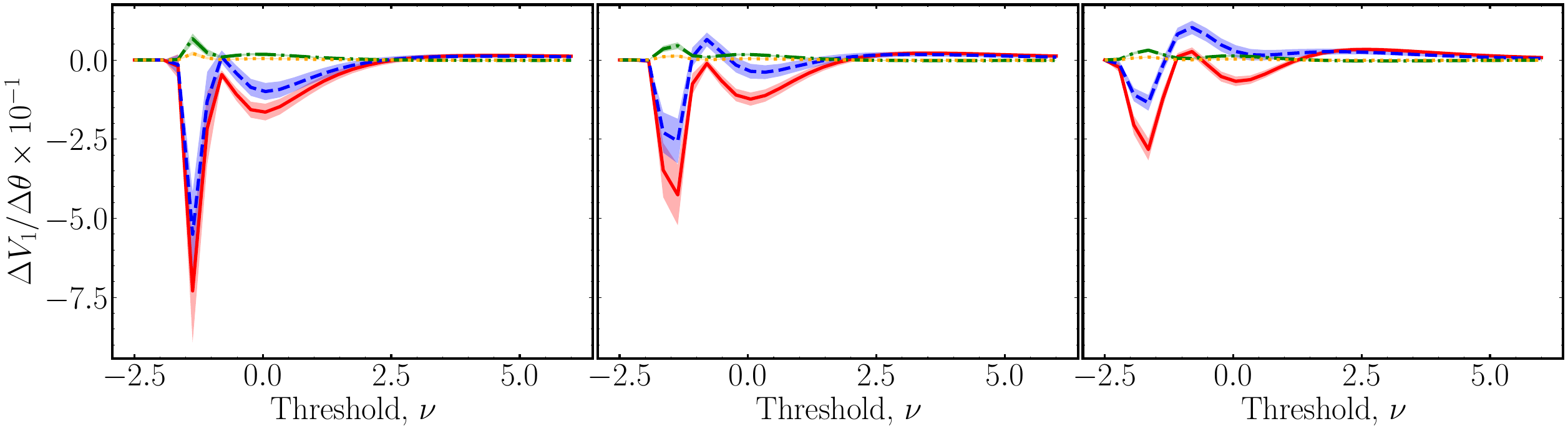}\\
    \vspace{-0.85cm}
  	\includegraphics[width=0.972\textwidth]{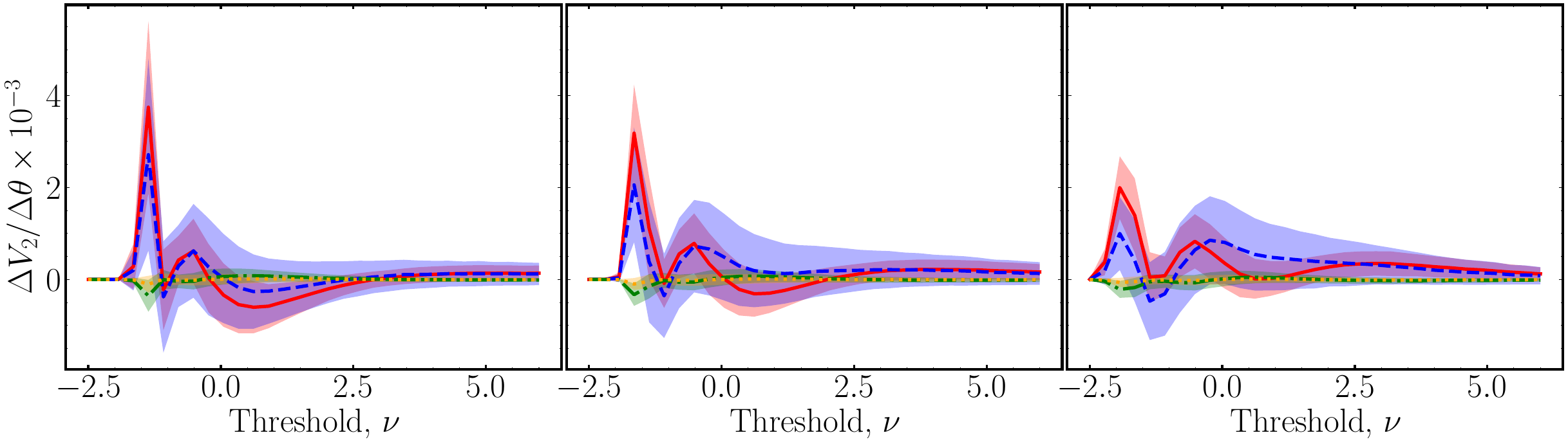}\\
  	\includegraphics[width=0.972\textwidth]{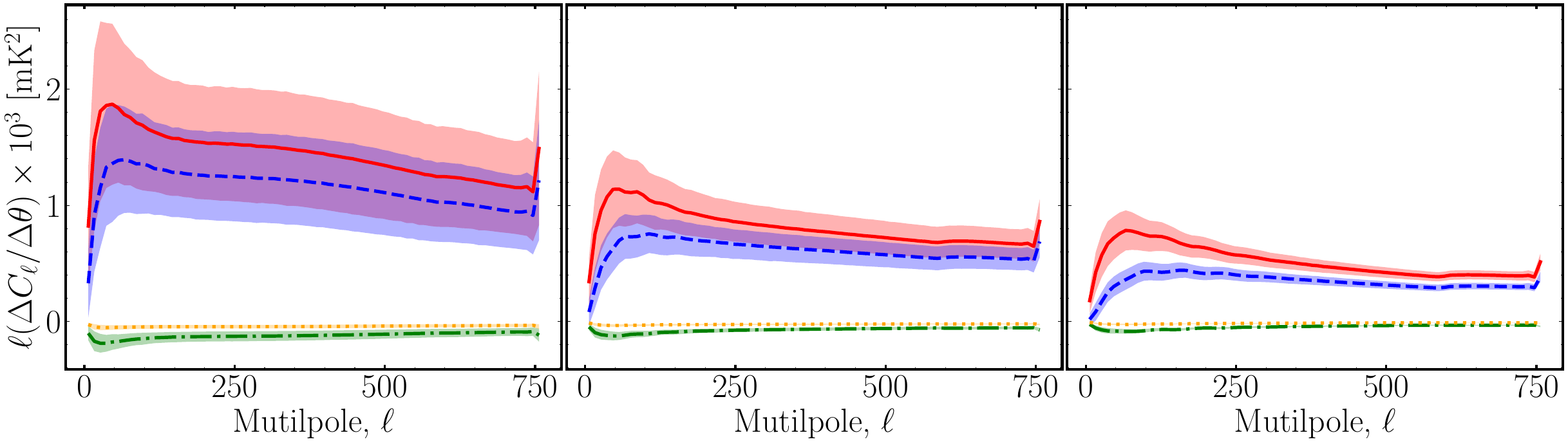}
    \caption{Average differences in the compressed summary statistics introduced by changing one cosmological parameter at a time with respect to the Planck 2018 cosmology \citep{2020/aghanim-planck}. From top to bottom we show the curves for the three MFs, Area ($V_0$), Perimeter ($V_1$), and Genus ($V_2$), and for the $C_\ell$, respectively.  The differences are calculated as $\Delta\,X \equiv \langle X - X^{\it fid} \rangle / \Delta\theta$, for $X = V_0, V_1, V_2$, and $C_\ell$, where $\Delta\theta = \theta - \theta^{\it fid}$ is the variation in the cosmological parameter value with respect to the fiducial value (1\%; see text for details). The coloured regions show the respective 1$\sigma$ from 1000 clean 21\,cm simulations (accounting for the 40 arcmin beam and the BINGO sky coverage). From left to right, the columns show results from the first, middle, and last compressed redshift ranges, as indicated in each plot of the first row. 
    }
    \label{fig:dif_MFs}
\end{figure*}

To evaluate the sensitivity of the MFs with the four cosmological parameters predicted here, we generate a set of 1000 clean 21\,cm mocks, with the BINGO beam and sky coverage, for the fiducial Planck 2018 cosmology \citep{2020/aghanim-planck}. 
Changing each of the four cosmological parameters one at a time, increasing it by 1\% from its fiducial value, we generate the other four sets of 1000 mocks each. 
These sets are generated using the same random seed used when generating the 1000 mocks for the fiducial cosmology. 
Then, we obtain the MFs and APS from each of the 30 redshift bins composing each mock, compress them into 5 bins (see discussion in Section \ref{sec:sensitivity_zbins}) and calculate the difference between the Area ($V_0$), Perimeter ($V_1$), Genus ($V_2$), and $C_\ell$ from the four sets of simulations and those from the fiducial cosmology, $V_k - V_k^{\it fid}$. 
The average difference $\Delta V_k = \langle V_k - V_k^{\it fid} \rangle / \Delta \theta$ from the 1000 mocks, scaled by the difference between the cosmological parameter $\theta$ and its fiducial value $\theta^{\it fid}$, as well as the respective 1$\sigma$ regions, for each cosmological parameter are shown in Figure \ref{fig:dif_MFs}. 
There we present the results for the first, middle, and last compressed (effective) redshift ranges.


\bsp	
\label{lastpage}
\end{document}